\documentclass[useAMS,usenatbib,usegraphicx]{mn2e}
\usepackage{longtable}
\usepackage{amsmath}
\usepackage{amssymb}
\usepackage{longtable}

\usepackage{lscape}

\usepackage{graphics}

\usepackage{color}

\RequirePackage{lineno}

\def\OII{[O\,{\sc ii}]}

\def\gs{\mathrel{\raise0.35ex\hbox{$\scriptstyle >$}\kern-0.6em
\lower0.40ex\hbox{{$\scriptstyle \sim$}}}}
\def\ls{\mathrel{\raise0.35ex\hbox{$\scriptstyle <$}\kern-0.6em
\lower0.40ex\hbox{{$\scriptstyle \sim$}}}}
\def\m@th{\mathsurround=0pt }
\def\eqalign#1{\null\,\vcenter{\openup1\jot \m@th
 \ialign{\strut\hfil$\displaystyle{##}$&$\displaystyle{{}##}$\hfil
 \crcr#1\crcr}}\,}

\long\def\symbolfootnote[#1]#2{\begingroup%
  \def\thefootnote{\fnsymbol{footnote}}\footnote[#1]{#2}\endgroup}

\title[Herschel reveals the obscured star formation in HiZELS H$\alpha$
        emitters at $z=1.47$]
      {Herschel reveals the obscured star formation in HiZELS H$\alpha$
        emitters at $z=1.47$}  
\author[E.~Ibar et al.]
{\parbox{\textwidth}{\raggedright E.~Ibar,$^{1,2}$\thanks{E-mail: \texttt{ibar@astro.puc.cl}}
D.~Sobral,$^{3}$
P.N.~Best,$^{4}$
R.J.~Ivison,$^{1,4}$
I.~Smail,$^{5}$
V.~Arumugam,$^{4}$
S.~Berta,$^{6}$
M.~B{\'e}thermin,$^{7,8}$
J.~Bock,$^{9,10}$
A.~Cava,$^{11}$
A.~Conley,$^{12}$
D.~Farrah,$^{13}$
E.~Le Floc'h,$^{7}$
D.~Lutz,$^{6}$
G.~Magdis,$^{7}$
B.~Magnelli,$^{6}$
S.~Ikarashi,$^{14}$
K.~Kohno,$^{14,15}$
G.~Marsden,$^{16}$
S.J.~Oliver,$^{13}$
M.J.~Page,$^{17}$
F.~Pozzi,$^{18}$
L.~Riguccini,$^{7}$
B.~Schulz,$^{9,19}$
N.~Seymour,$^{20,17}$
A.J.~Smith,$^{13}$
M.~Symeonidis,$^{17}$
L.~Wang,$^{13,5}$
J.~Wardlow$^{21}$ and
M.~Zemcov$^{9,10}$}\vspace{0.4cm}\\
\parbox{\textwidth}{\raggedright $^{1}$UK Astronomy Technology Centre, Science and Technology Facilities Council, Royal Observatory, Blackford Hill, Edinburgh EH9 3HJ, UK\\
$^{2}$Instituto de Astrof\'isica, Facultad de F\'isica, Pontificia Universidad Cat\'olica de Chile, Casilla 306, Santiago 22, Chile
\\
$^{3}$Leiden Observatory, Leiden University, P.O.\ Box 9513, NL-2300 RA Leiden, The Netherlands\\
$^{4}$Institute for Astronomy, University of Edinburgh, Royal Observatory, Blackford Hill, Edinburgh EH9 3HJ, UK\\
$^{5}$Institute for Computational Cosmology, Durham University, South Road, Durham DH1 3LE, UK\\
$^{6}$Max-Planck-Institut f\"ur Extraterrestrische Physik (MPE), Postfach 1312, 85741, Garching, Germany\\
$^{7}$Laboratoire AIM-Paris-Saclay, CEA/DSM/Irfu - CNRS - Universit\'e Paris Diderot, CE-Saclay, pt courrier 131, F-91191 Gif-sur-Yvette, France\\
$^{8}$Institut d'Astrophysique Spatiale (IAS), b\^atiment 121, Universit\'e Paris-Sud 11 and CNRS (UMR 8617), 91405 Orsay, France\\
$^{9}$California Institute of Technology, 1200 E. California Blvd., Pasadena, CA 91125, USA\\
$^{10}$Jet Propulsion Laboratory, 4800 Oak Grove Drive, Pasadena, CA 91109, USA\\
$^{11}$Departamento de Astrof\'isica, Facultad de CC. F\'isicas, Universidad Complutense de Madrid, E-28040 Madrid, Spain\\
$^{12}$Center for Astrophysics and Space Astronomy 389-UCB, University of Colorado, Boulder, CO 80309, USA\\
$^{13}$Astronomy Centre, Dept. of Physics \& Astronomy, University of Sussex, Brighton BN1 9QH, UK\\
$^{14}$Institute of Astronomy, University of Tokyo, 2-21-1 Osawa, Mitaka, Tokyo 181-0015, Japan\\
$^{15}$Research Center for the Early Universe, University of Tokyo, 7-3-1 Hongo, Bunkyo, Tokyo 113-0033, Japan\\
$^{16}$Department of Physics \& Astronomy, University of British Columbia, 6224 Agricultural Road, Vancouver, BC V6T~1Z1, Canada\\
$^{17}$Mullard Space Science Laboratory, University College London, Holmbury St. Mary, Dorking, Surrey RH5 6NT, UK\\
$^{18}$INAF-Osservatorio Astronomico di Roma, via di Frascati 33, 00040 Monte Porzio Catone, Italy\\
$^{19}$Infrared Processing and Analysis Center, MS 100-22, California Institute of Technology, JPL, Pasadena, CA 91125, USA\\
$^{20}$CSIRO Astronomy \& Space Science, PO Box 76, Epping, NSW 1710, Australia\\
$^{21}$Dept. of Physics \& Astronomy, University of California, Irvine, CA 92697, USA}}
\begin{document}

\date{Accepted 2013 July 4. Received 2013 July 3; in original form 2012 November 30}

\pagerange{\pageref{firstpage}--\pageref{lastpage}} \pubyear{2012}


\label{firstpage}

\maketitle

\clearpage

\begin{abstract}
 We describe the far-infrared (far-IR; rest-frame
  8--1000\,$\mu$m) properties of a sample of 443 H$\alpha$-selected
  star-forming galaxies in the COSMOS and UDS fields detected by the
  HiZELS imaging survey. Sources are identified using narrow-band
  filters in combination with broad-band photometry to uniformly
  select H$\alpha$ (and \OII\ if available) emitters in a narrow
  redshift slice at $z=1.47\pm0.02$. We use a stacking approach in
  {\it Spitzer}-MIPS mid-IR, {\it Herschel}-PACS/SPIRE far-IR (from
  the PACS Evolutionary Probe (PEP) and {\it Herschel} Multi-tiered
  Extragalactic Surveys (HerMES) surveys) and AzTEC mm-wave images to
  describe their typical far-IR properties. We find that HiZELS
  galaxies with observed H$\alpha$ luminosities of $L({\rm
    H\alpha})_{\rm obs}\approx 10^{8.1-9.1}\,$L$_\odot$
  ($\approx\,10^{41.7-42.7}$\,erg\,s$^{-1}$) have bolometric far-IR
  luminosities of typical luminous IR galaxies, $L(8-1000\,\mu{\rm
    m})\approx 10^{11.48^{+0.04}_{-0.05}}$\,L$_\odot$. Combining the
  H$\alpha$ and far-IR luminosities, we derive median star-formation
  rates of SFR$_{\rm H\alpha,FIR}=32\pm5$\,M$_\odot$\,yr$^{-1}$ and
  H$\alpha$ extinctions of $A_{\rm H\alpha}=1.0\pm0.2$\,mag. Perhaps
  surprisingly, little difference is seen in typical HiZELS extinction
  levels compared to local star-forming galaxies. We confirm previous
  empirical M$_\star$-$A_{\rm H\alpha}$ relations and the little or no
  evolution up to $z=1.47$. For HiZELS galaxies (or similar samples)
  we provide an empirical parametrisation of the SFR as a function of
  rest-frame $(u-z)$ colours and 3.6\,$\mu$m photometry -- a useful
  proxy to aid in the absence of far-IR detections in high-$z$
  galaxies.  We find that the observed H$\alpha$ luminosity is a
  dominant SFR tracer when rest-frame $(u-z)$ colours are
  $\lesssim\,0.9$\,mag or when {\it Spitzer}-3.6\,$\mu$m photometry is
  fainter than 22\,mag (Vega) or stellar masses are lower than
  10$^{9.7}\,$M$_\odot$. We do not find any correlation between the
  \OII/H$\alpha$ and far-IR luminosity, suggesting that this emission
  line ratio does not trace the extinction of the most obscured
  star-forming regions, especially in massive galaxies where these
  dominate. The luminosity-limited HiZELS sample tends to lie above of
  the so-called `main sequence' for star-forming galaxies, especially
  at low stellar masses, indicating high star-formation efficiencies
  in these galaxies. This work has implications for SFR indicators and
  suggests that obscured star formation is linked to the assembly of
  stellar mass, with deeper potential wells in massive galaxies
  providing dense, heavily obscured environments in which stars can
  form rapidly.  
\end{abstract}


\section{Introduction}
\label{introsec}

Historically, the classical star-formation rate (SFR) indicator has
been H$\alpha$ ($\lambda_{\rm rest}=656.3$\,nm) luminosity -- a
well-calibrated probe of instantaneous emission from massive, young
stars ($<20$\,Myr and $>8$\,M$_\odot$;
e.g.\ \citealt{Kennicutt98}). To extrapolate the observed starlight
from O and B stars to the total SFR requires careful consideration of
the initial mass function (IMF) of the stellar population and the
amount of extinction (scattering and absorption by dust) suffered by
the starlight. Discrepancies of up to $\sim$30\,per~cent can be found
amongst previously published H$\alpha$-based SFR calibrations, due
mainly to the use of different models of stellar evolution and stellar
atmospheres.

Of all the assumptions required to convert from observed quantities
into SFRs, the main limitation is the sensitivity of H$\alpha$ flux to
extinction ($A_{\rm H\alpha}$, where the intrinsic H$\alpha$
luminosity is defined as $L({\rm H\alpha})_{\rm int} = L({\rm
  H\alpha})_{\rm obs}\times 10^{0.4\,A_{\rm H\alpha}}$). The observed
line fluxes represent only a fraction of the intrinsic emission, with
typical values of $A_{\rm H\alpha}$ found to be $\approx 0.8-1.1$\,mag
in optically-selected samples \citep{Niklas97, Sobral12}. In general,
when the level of extinction is low or moderate, $A_{\rm H\alpha}\ls
3$\,mag, the difference between the observed Balmer decrement
(H$\alpha$/H$\beta$; e.g.\ see \citealt{Calzetti01}) and the
theoretical expectation (2.86 for Case B recombination: electron
temperature $T_{\rm e}=10^4$\,K and density $n_{\rm e}=10^2\,{\rm
  cm}^{-3}$; \citealt{Brocklehurst71, Kennicutt98}) can be used to
determine the amount of extinction (assuming a model for the
wavelength-dependence of the attenuation e.g.\ \citealt{Fischera05})
as this H$\alpha$/H$\beta$ ratio scales directly with the total
ionising flux of the embedded stars. Unfortunately, the combination of
extinction and increasing redshift make detection of Hydrogen lines
difficult (especially for H$\beta$, $\lambda_{\rm
  rest}=486.1$\,nm). Indeed, to study star-formation processes via
conventional means at high redshift is a major challenge
(e.g.\ \citealt{Stott13}).

A way to measure the SFR at high redshift is by considering that the
UV/optical photons that are absorbed by the surrounding media are
re-emitted in the far-IR waveband (e.g.\ \citealt{Heinis13}), meaning
that far-IR observables can be used as a tracer of the obscured SFR
and/or the amount of extinction in a galaxy. For example, if all the
starlight is absorbed then the system works as a calorimeter and the
far-IR becomes the ideal tracer of SFR (\citealt{Lacki10}). This
measure includes those contributions from heavily obscured star
formation and those from old stellar populations
\citep[e.g.][]{Salim09}. These far-IR SFR estimates have an intimate
relationship with the level of extinction suffered by the
starlight. We stress that we think of extinction as an average
quantity measured towards all star-forming regions of a galaxy, and it
therefore presents several other intricate dependencies, e.g.\ on
geometry, luminosity, mass, environment, radiation fields, etc.
\citep[e.g.][]{Dutton10}.

A previous study in the local Universe, $\left<z\right>\approx0.08$,
by \citet{bGarn10}, using data from the Sloan Digital Sky Survey
(SDSS), showed that one of the strongest parameters correlating with
the level of extinction in galaxies is stellar mass, $M_\star$
\citep[see also][]{Brinchmann04, Gilbank10, Wuyts11}. 
These studies suggest that the level of extinction produced by the
material in and surrounding their star-forming regions increase as the
galaxies built up their stellar mass. Using H$\alpha$- and $H$-band
selected galaxies, \citet{Sobral12} and \citet{Hilton12} find that
this behaviour seems to hold even at $z\approx 1.5-3$. On the other
hand, various studies have shown that mass plays a key role in driving
the amount of star formation \citep[e.g.\ at $z\sim 1.5$,
  see][]{Noeske07, Elbaz07, Daddi07, Pannella09}, where the SFR is
found to be roughly linearly correlated to the stellar mass and
defines a typical value for the specific SFR
(sSFR\,$=$\,SFR/$M_\star$), where more violent star formation is seen
in more massive galaxies.

Using data from the {\it Herschel Space Observatory}\footnote{{\it
    Herschel} is a ESA space observatory with science instruments
  provided by the European-led Principal Investigator consortia and
  with important participation from NASA.} \citep{Pilbratt10},
\citet{Elbaz11} propose, somewhat controversially, the existence of
two modes of star formation: `normal' galaxies which lie in a
well-defined parameter space (the `main sequence') defined in a plot
of sSFR versus redshift, and `starburst' galaxies which present an
excess in sSFR related to an increment of efficiency in compact
star-forming regions probably triggered by the merger of two or more
galaxies \citep[e.g.][]{Daddi10}. The controversy comes from the fact
that these results are sensitive to the way by which `star-forming
galaxies' are selected \citep{Sobral11, Karim11} as the sSFR can
change as more passive galaxies satisfy the applied selection criteria
(Cirasuolo et al., in prep).

To explore the intimate relationship between SFR, $M_\star$ and
$A_{\rm H\alpha}$, we make use of $\sim 2$\,deg$^2$ imaged by the
High-redshift Emission Line Survey\footnote{For more details on the
  survey, progress and data release, see
  http://www.roe.ac.uk/ifa/HiZELS} (HiZELS; \citealt{Geach08,
  Sobral09, Sobral13}) in the cosmic evolution survey (COSMOS,
$\sim$\,1.45\,sq.\,deg.; \citealt{Scoville07c}) and the UKIRT Infrared
Deep Sky Survey (UKIDSS; \citealt{Lawrence07}) Ultra Deep Survey (UDS,
$\sim$\,0.67\,sq.\,deg.; Almaini et al., in prep) fields. We extract a
large and unique sample of relatively low-luminosity ($L({\rm
  H\alpha})_{\rm obs}\approx 10^{41.5-42.5}\,$erg\,s$^{-1}$)
star-forming galaxies, using a tuned narrow-band-filter technique to
pick up large numbers of simultaneous H$\alpha$ and \OII\ (if
available) emitters (alleviating the need for spectroscopic redshifts)
at a well-defined $z=1.47$ redshift (\citealt{Sobral12}).

Taking advantage of the plethora of multi-wavelength coverage in the
UDS and COSMOS fields, we describe the far-IR (rest-frame
8--1000\,$\mu$m) properties of the HiZELS sample using data taken by:
{\it Spitzer}-Multiband Imaging Photometer (MIPS; \citealt{Rieke04})
at 24 and 70\,$\mu$m; {\it Herschel} Photodetector Array Camera and
Spectrometer (PACS; \citealt{Poglitsch10}) at 100 and 160\,$\mu$m as
part of the PACS Evolutionary Probe (PEP; \citealt{Lutz11}) survey;
{\it Herschel} Spectral and Photometric Imaging Receiver (SPIRE;
\citealt{Griffin10}) at 250, 350 and 500\,$\mu$m as part of the {\it
  Herschel} Multi-tiered Extragalactic Survey (HerMES$^\dagger$;
\citealt{Oliver12}); and the Astronomical Thermal Emission Camera
(AzTEC; \citealt{Wilson08}) at 1100\,$\mu$m while mounted at the James
Clerk Maxwell Telescope (JCMT) and at the Atacama Submillimeter
Telescope Experiment (ASTE) -- more details about these data are given
in Table~\ref{table1}. Similar works were done for a sample at
$z=2.23$ by \citet{Geach08} using {\it Spitzer} at 70 and 160\,$\mu$m,
achieving upper limits near the peak of the SED, and at $z=0.84$ using
24-$\mu$m imaging \citep{aGarn10}.

Direct H$\alpha$ measurements, in combination with {\it Spitzer}, {\it
  Herschel} and AzTEC imaging provide an ideal framework for a
detailed description of the star-formation activity at $z=1.47$ (near
the peak of cosmic star-formation history, where most of the galaxy
mass was assembled; e.g.\ \citealt{Dickinson03}), and its dependencies
on parameters such as luminosity, stellar mass, \OII/H$\alpha$ ratio
and rest-frame colours. We mainly make use of a recent parametrisation
of the SFR (\citealt{Kennicutt09}) based on a linear combination of
the observed H$\alpha$ and bolometric far-IR (rest-frame
8--1000-$\mu$m) luminosities. This estimate is suitable for both
far-IR- and optically-selected star-forming galaxies.

In this paper, the sample is described in \S\ref{data_sample} and our
analysis of the stacked far-IR measurements is explained in
\S\ref{stacking_section}. The results are discussed in
\S\ref{discussion_section} and our conclusions are summarised in
\S\ref{conclusion_section}. Throughout the text, we adopt a Salpeter
IMF \citep{Salpeter55} and estimate the contribution from the
thermally-pulsing asymptotic giant branch (TP-AGB;
e.g.\ \citealt{Trujillo07}) in our derived stellar masses. We use a
$\Lambda$CDM cosmology with $H_0=70$\,km\,s$^{-1}$\,Mpc$^{-1}$,
$\Omega_{\rm M}=0.3$ and $\Omega_{\Lambda}=0.7$.

%
%
%

\begin{table*}
\centering
\caption{Broad-band data used in this work. Noise (r.m.s.) values are
  obtained from the pixel fluctuation seen in the whole area used for
  stacking, including the normalisation $\eta$ found between fitted
  peaks and catalogued fluxes (see
  \S\,\ref{empirical_cal_section}). Note these values are slightly
  different with respect to published ones. The references for each of
  the catalogues and images are as follow: {\bf UDS:} 24\,$\mu$m, from
  the {\it Spitzer} UKIDSS UDS (SpUDS; data version delivery S18.7
  from the NASA/IPAC Infrared Science Archive$^\S$); 70\,$\mu$m, from
  the {\it Spitzer} Wide-area Infrared Extragalactic Survey (SWIRE;
  \citealt{Surace05}, data version delivery$^\S$ DR3-S11); {\it
    Herschel} PACS 100 and 160\,$\mu$m data from the HerMES survey
  (UDS deep level-3 field; reduced as in
  \citealt{Ibar10} but using an improved deglitching approach); SPIRE
  250, 350 and 500\,$\mu$m maps (SMAP\_v4.2; \citealt{Levenson10}) and
  catalogues (SCAT\_SXT\_iDR1; \citealt{Smith12}) retrieved from the
  {\it Herschel} Database in Marseille (HeDaM$^\ast$) of the HerMES
  survey; AzTEC 1100\,$\mu$m data taken at JCMT
  \citep{Austermann10}. {\bf COSMOS:} 24 and 70\,$\mu$m, from the {\it
    Spitzer} coverage of the COSMOS field (S-COSMOS; GO2+GO3a+GO3b
  Delivery v1 from the NASA/IPAC Infrared Science
    Archive$^\ddagger$; \citealt{Sanders07,LeFloch09,Frayer09}); PACS
  100 and 160\,$\mu$m from the PEP survey; SPIRE 250, 350 and
  500\,$\mu$m from HerMES (SMAP\_v4.2 and SCAT\_SXT\_iDR1 from HeDaM);
  AzTEC 1100\,$\mu$m data taken at ASTE (\citealt{Scott08}).}
\begin{tabular}{|cccccc|}
\hline
Telescope/&
Central $\lambda$ &
{\sc fwhm} &
pixelsize &
r.m.s.\ [UDS]&
r.m.s.\ [COSMOS] 
\\
detector &
($\mu$m) &
(arcsec) &
(arcsec) &
(mJy\,beam$^{-1}$) &
(mJy\,beam$^{-1}$) 
\\
\hline
{\it Spitzer}-MIPS &
24 &
6.0 &
1.2 &
0.025 &
0.022
\\
{\it Spitzer}-MIPS &
70 &
18.2   &
4.0 &
3.7 &
2.8
\\
{\it Herschel}-PACS &
100 &
7.03  &
2.0 &
2.1 &
1.9
\\
{\it Herschel}-PACS &
160 &
11.55 &
3.0 &
4.7 &
4.3
\\
{\it Herschel}-SPIRE &
250 &
18.15  &
6.0 &
5.5 &
4.5
\\
{\it Herschel}-SPIRE &
350 &
25.15  &
8.33 &
6.3 &
5.3
\\
{\it Herschel}-SPIRE &
500 &
36.30 &
12.0 &
7.1 &
5.8
\\
JCMT-AzTEC &
1100 &
18.0 &
3.0 &
1.4 &
\\
ASTE-AzTEC &
1100 &
30.0 &
3.0 &
 &
1.4
\\
\hline
\end{tabular}
\label{table1}
\end{table*}

\section{The Sample of H$\mathbf{\alpha}$ emitters at $\mathbf{z=1.47}$}
\label{data_sample}

HiZELS uses narrow-band filters to detect H$\alpha$ emission at a
variety of redshifts, up to $z=2.23$ (\citealt{Sobral13}). Given the
nature of the H$\alpha$ emission, HiZELS selects only young
star-forming galaxies and AGN. Distinguishing between H$\alpha$ and
any other emission lines at other redshift is a critical
step. Double-matched narrow-band surveys detecting strong emission
lines offer a good way of mitigating this problem
(e.g.\ \citealt{Sobral12}). For star-forming galaxies at $z\sim 1.5$
we make use of the fact that the H$\alpha$ line is detectable in the
$H$ band while the \OII-372.7-nm emission can be observed at the red
end of the $z'$ band. As shown in \citet{Sobral12}, by combining deep
broad-band photometry with tuned double-narrow-band imaging -- where
the NB921 narrow-band filter on Subaru/Suprime-Cam detects \OII\ and
the NB$_H$ filter on UKIRT/WFCAM detects H$\alpha$ at the same
redshift -- it is possible to conduct an effective survey of
line-emitting sources at $z=1.47$.

We extract HiZELS samples from \citet{Sobral12, Sobral13} which
provide uniform H$\alpha$ coverage across the UDS, reaching an average
effective flux limit (3\,$\sigma$) of $S_{\rm
  H\alpha}$\,$\approx$\,$9\times10^{-17}$\,erg\,s$^{-1}$\,cm$^{-2}$,
while the matched Subaru (\OII) narrow-band survey reaches an
effective flux of $S_{\rm [OII]}$\,$\approx$\,$ 9\times
10^{-18}$\,erg\,s$^{-1}$\,cm$^{-2}$. The depth of the NB921 imaging
provides counterparts for all H$\alpha$ emitters in the UDS and
therefore allows a clean selection of the H$\alpha$ sample. In COSMOS,
the H$\alpha$ narrow-band imaging was intentionally designed to obtain
a `wedding cake' survey (i.e.\ deeper than UDS in small regions),
while the \OII\ imaging has relatively uniform coverage, resulting in
a large number of COSMOS H$\alpha$ emitters without a \OII\ detection
($\sim$\,50\,per~cent). For a cleaner selection of H$\alpha$ emitters
and not other emission line objects, we make use of the available
broadband photometry in these fields. As described in
\citet{Sobral12}, we use colour-colour criteria (a method similar to
the {\it BzK} diagnostic; \citealt{Daddi04}) in a $B-R$ vs.\ $i-K$
diagram to first remove low-redshift contaminants, and then in $i-z$
vs.\ $z-K$ to remove the high-redshift emitters (e.g.\ [O\,{\sc iii}]
and H$\beta$). From all H$\alpha$ candidates, these colour-colour
criteria and the matched \OII\ detection remove $\sim\,50\,$per~cent
of the candidates, resulting in a tight redshift distribution ($\Delta
z\approx 0.02$) as evidenced by the small number of sources with
available spectroscopic redshifts ($\sim 5$\,\% of the sample). The
total number of H$\alpha$ emitters identified by HiZELS at
$z=1.47\pm0.02$ is 188 in UDS and 325 in COSMOS fields. 

\citet{Sobral13} estimate that after applying the mentioned
colour-colour criteria to a sample which does not present
NB921-\OII\ detections, the level of contamination of emitting
galaxies at different redshifts is of the order of $\sim$\,15\%. Some
of these contaminants could come from H$\beta$, [O\,{\sc iii}] or
\OII\ at $z=2.2$ and $z=3.3$, or possibly from galaxies at $z<1$ over
a wide range of possible emission lines.  Given that $\sim$\,50\% of
COSMOS galaxies lack \OII\ detections due to no data being available,
or because it is too shallow, we expect that the overall contamination
in the COSMOS sample should be of the order of $\sim$\,7.5\%,
i.e.\ $\sim$\,5\% in the full UDS+COSMOS sample. We do not expect
these contaminants to distribute in well defined redshifts, so their
contribution to the stacks is unknown. We assume that a possible
contamination of 5\% will not be sufficient to significantly modify
the median stacks of our analysis.

\phantom{.}~\symbolfootnote[0]{$^\dagger$\scriptsize http://hermes.sussex.ac.uk}
\phantom{.}~\symbolfootnote[0]{$^\S$\scriptsize http://irsa.ipac.caltech.edu/data/SPITZER/docs/spitzermission/observingprograms}
\phantom{.}~\symbolfootnote[0]{$^\ast$\scriptsize http://hedam.oamp.fr/HerMES/index.php} 
\phantom{.}~\symbolfootnote[0]{$^\ddagger$\scriptsize http://irsa.ipac.caltech.edu/Missions/spitzer.html}

\begin{figure*}
   \centering
   \includegraphics[scale=0.55]{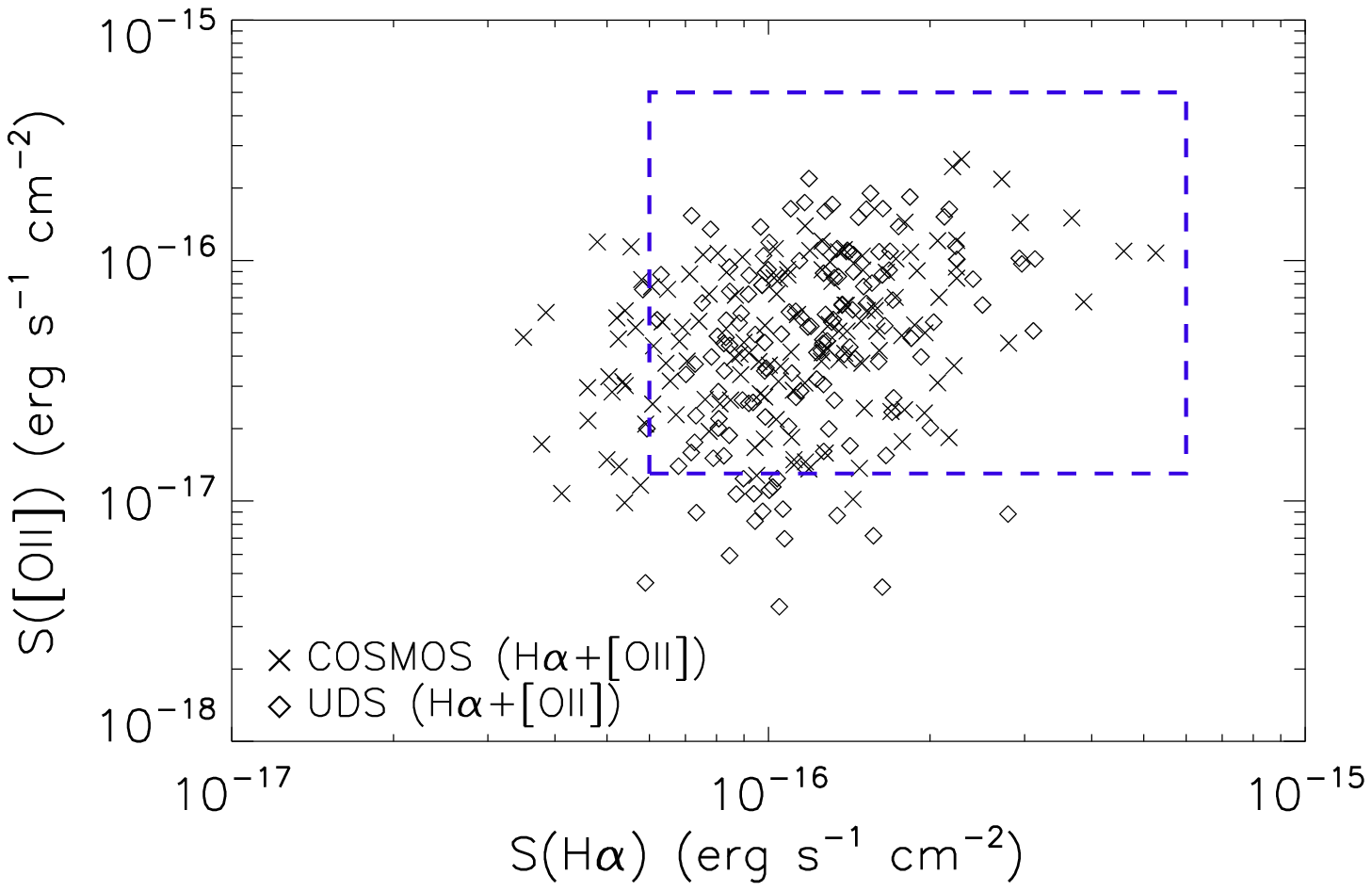}
   \includegraphics[scale=0.55]{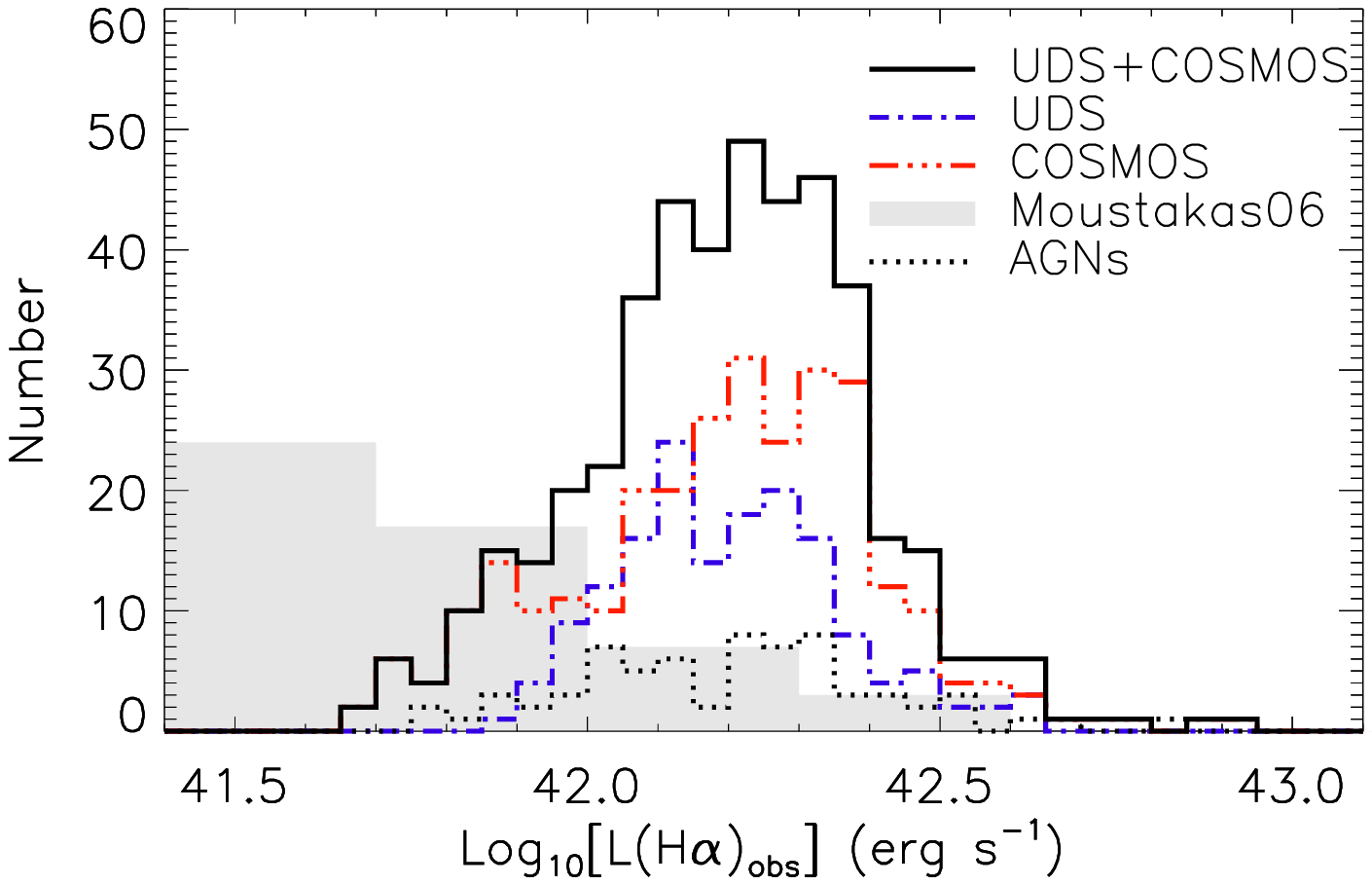}
   \caption{{\bf Left:} Measured narrow-band H$\alpha$ and \OII\ fluxes
     for sources in the UDS and COSMOS fields. The dashed rectangle
     shows the region used to show the consistency of stacked signals
     from both fields (see Fig.~\ref{merge_stacks}, {\it left}). {\bf
       Right:} Distribution of observed H$\alpha$ luminosities
     (corrected for [N\,{\sc ii}] contamination but not for
     extinction). In black (thick solid), blue (dot-dashed) and red
     (3$\times$dot-dashed) lines: the merged, UDS and COSMOS samples,
     respectively. Black dotted line shows the identified AGN
     distribution as described in \S\,\ref{AGN_removal_section}.  The
     shaded area corresponds to the bright end of the source
     distribution used to calibrate the \citet{Kennicutt09} SFR
     correlations in the local Universe (\citealt{Moustakas06}).}
   \label{sample_figure}
\end{figure*}

In Fig.~\ref{sample_figure} we show the observed H$\alpha$ luminosity
distribution ($L({\rm H}\alpha)_{\rm obs}$, not corrected for
extinction) for the HiZELS samples. Values of $L({\rm H}\alpha)_{\rm
  obs}$ have been corrected by removing the flux estimated to be
contributed by the adjacent [N\,{\sc ii}] doublet at 654.8 and
658.3\,nm, following \citet{Villar08}, as presented in
\citet{Sobral12}. Based on a recent estimation of the
  point-spread-function in the narrow band images, the H$\alpha$ and
  \OII\ photometry increases by $\sim$\,30\,per cent with respect to
  those presented in \citet{Sobral12, Sobral13}. This is an aperture
  correction factor introduced to take into account the flux missed
  at $>$\,2\,arcsec radius. The observed luminosity distribution can
be roughly characterised by $L({\rm H\alpha})_{\rm obs}=10^{42.2\pm
  0.2}\,$erg\,s$^{-1}$, i.e.\ to an equivalent SFR$_{\rm
  H\alpha}\approx 32$\,M$_{\odot}$\,yr$^{-1}$ (\citealt{Kennicutt98}),
assuming an average extinction of 1\,mag for the H$\alpha$
luminosities. We note that these observed luminosities are within the
range of those which define the SFR correlations in the local Universe
(see Fig.~\ref{sample_figure}; \citealt{Kennicutt09} using sources
from \citealt{Moustakas06}), although we are inevitably biased against
extreme extinction (undetected at H$\alpha$; $A_{\rm H\alpha}\gs 3$)
and towards galaxies with high SFR and young stellar populations.

%

\subsection{Removing AGN from the sample}
\label{AGN_removal_section}

We have deliberately chosen a conservative approach to account for AGN
contamination. We have removed all sources previously catalogued at
X-ray wavelengths (\citealt{Ueda08}; \citealt{Cappelluti09}) as this
$\sim$\,2\,keV emission at $z=1.47$ is expected to be produced via
inverse Compton scattering by a compact and highly ionised region
surrounding an AGN. We might remove some sources presenting powerful
thermal X-ray emission, although this should not affect our analysis
as only 6 and 3 sources are X-ray emitters in the UDS and COSMOS
fields, respectively. We have also used available 1.4-GHz images
(\citealt{Schinnerer10}; Arumugam et al., in prep) to identify
synchrotron emission produced by an AGN. Assuming a typical SFR$_{\rm
  H\alpha}\approx 32$\,M$_{\odot}$\,yr$^{-1}$ for the HiZELS
population (based on $A_{\rm H\alpha}=1$) and the validity of the
far-IR/radio correlation at high redshift
(\citealt{Ibar08,aIvison10}), we remove all sources having 1.4-GHz
flux densities larger than 150\,$\mu$Jy given that their expected
radio luminosities would put them $\gs$\,5-$\sigma$ away from the
far-IR/radio correlation. In addition, similar to \citet{aGarn10} we
use template SEDs of star-forming galaxies and AGN to fit the
multi-wavelength broad-band photometry available for our sources (see
\S\,\ref{stellar_mass_section}). All those best-fitted as AGN are
identified with red rest-frame mid-IR colours. We average the observed
{\it Spitzer} 3.6- and 4.5-$\mu$m fluxes to estimate the rest-frame
1.6-$\mu$m flux density, and the 5.8- and 8.0-\,$\mu$m fluxes for the
2.8-$\mu$m rest-frame flux density. All SED-fitted AGN, the bulk of
the X-ray sources and those sources spectroscopically classified as
AGN using the BPT diagram (\citealt{Stott13}, Sobral et al.\, in
prep.) are identified to have a rest-frame flux density ratio
2.8\,$\mu$m/1.6\,$\mu$m\,$>$\,1. We apply this simple threshold to
remove further potential AGN from the sample. Note that this near-IR
criterion is the same that makes \citet{Lacy04} and \citet{Stern05}
methods work, but we optimise it for $z=1.47$ galaxies in order to
minimise the errors in IRAC photometry.

These criteria (X-ray/radio/mid-IR) identify a total of 30 (6/9/22)
and 40 (3/5/35) potential AGN in the UDS and COSMOS fields,
respectively. Some of them are identified by more than one
criterion. Our final star-forming galaxy sample consist of 443 sources
(158 in UDS and 285 in COSMOS) at a well defined redshift, $z=1.47$.

\section{The far-IR properties of HiZELS galaxies}
\label{stacking_section}

We describe the far-IR properties of the HiZELS sample by taking
advantage of the plethora of multi-wavelength coverage in the UDS and
COSMOS fields. We note, however, that out of 443 selected star-forming
galaxies, only 10 (2\,per~cent) of them have catalogued {\it
  Herschel}/SPIRE-250\,$\mu$m sources (\citealt{Smith12}) within
2\,arcsec -- near the peak of the far-IR SED. They all
have $S_{\rm 250\mu m}<40\,$mJy at a significance of
$\ls$\,5\,$\sigma$ and there is no particular trend for 250\,$\mu$m
fluxes with observed H$\alpha$ luminosity. In contrast, within the
possible AGN population there are 5 detections (out of 70,
i.e.\ 7\,per~cent) at 250\,$\mu$m suggesting typically brighter far-IR
luminosities for this population. These small number of detections,
however, are not sufficient to provide a robust view to the HiZELS
population as a whole.

In this paper we use a stacking analysis to tackle the far-IR
properties of the H$\alpha$ galaxies. Stacking is a statistical method
which consists of cutting out a significant number of map regions
centred at the position of known sources (e.g.\ see details at
\citealt{Kurczynski10, bBethermin12, Heinis13}). When all these maps
are averaged together (pixel-by-pixel), signals at the image's centre
can emerge from the noise. These signals represent averaged (or
median) properties for the stacked population. The reliability of this
approach highly depends on the common nature of the parent population,
where statistical quantities are robust.

\subsection{Stacking and flux density measurements}
\label{SED_stack_section}

In this work, all images we use for stacking (at 24, 70,
  100, 160, 250, 350, 500 and 1100\,$\mu$m) have resolutions ({\sc
  fwhm}) much larger than the sub-arcsec astrometric uncertainties of
the HiZELS sample ($\sim$\,0.25\,arcsec), so for simplicity we
confidently assume the resulting stacked signals are
`point-like'. Images do not cover exactly the same sky
  area (see Table~\ref{table1}), so these differences
  imply that sources outside the coverage, or in noisy regions, are
  flagged differently for each map. The percentages of stacked sources
  per image are shown Table~\ref{table_cover}. 

We use arbitrary 91\,pixel\,$\times$\,91\,pixel\,$\times N$ (where $N$
is the number of stacked sources) data cubes for each waveband. Maps
extracted from the images ($S_i$) are centred at the closest map-pixel
to the source position. The data cubes are then collapsed by taking
the median signal in each map-pixel yielding simple $91\times
91$-pixel\,$^2$ images for each waveband. We prefer a
  median stack as this minimises the effect produced by outliers
  (e.g.\ by nearby bright galaxies) in the map-pixel
  distributions. In most cases, especially at 24, 250 and
350\,$\mu$m, a clear signal appears at the image centre
(results are shown in Table~\ref{table5}).

%
\begin{table}
\caption{The percentage of HiZELS galaxies (from a total
    of 158 in UDS and 285 in COSMOS) presenting imaging coverage at
    different wavelengths. Almost all images provide more than 90\%
    coverage, with the exemption of both AzTEC images which miss
    $\sim$\,20\% of the sources, and the deep {\it Herschel}-PACS UDS
    map which only covers 40\% of the sample (see PACS errors in
    Fig.~\ref{merge_stacks}).}  \centering
\begin{tabular}{|ccccccccc|}
\hline
 $\lambda$ &  UDS      & COSMOS \\ 
 ($\mu$m)  &  Cov(\%)  & Cov(\%) \\
\hline
 24   &  94 & 100 \\
 70   & 100 & 100 \\
 100  &  39 &  97 \\
 160  &  39 &  97 \\
 250  & 100 & 100 \\
 350  & 100 & 100 \\
 500  & 100 & 100 \\
 1100 &  79 &  78 \\
\hline
\end{tabular}
\label{table_cover}
\end{table}

In each of the stacked maps, we remove the median sky
background level ($B_{\rm MC}$). This level is estimated using a
Monte-Carlo simulation (100 realisations following the same approach
to create the stacks) randomising the source positions within
5\,arcmin from their original locations. This background subtraction
is found to be essential in order to properly co-add stacked signals
coming from different fields (see Eqn.~\ref{stack_eqn}).

\subsubsection{Spitzer stacks}

In {\it Spitzer} images (see Table~\ref{table1}), we use a 2D-Gaussian
fit ({\sc idl} routine {\sc mpfit2peak}) to extract the central
stacked peaks. The fit is performed using the following constraints:
the peak must be close to the central position ($\Delta{\rm
  R.A.},\Delta{\rm Dec.}$\,$<$\,{\sc fwhm}/2); the sky level is fixed
at zero (since the background has already been subtracted); the width
({\sc fwhm}) is fixed at the appropriate one for a point source (as
given in Table~\ref{table1}). From these Gaussian fits we extract the
peak value. Note that {\it Spitzer} images are in units of
MJy\,sr$^{-1}$, so this peak value needs a conversion factor to obtain
the integrated flux density (see below
\S\,\ref{empirical_cal_section}). In particular, we have arbitrarily
increased (by $3\times$) the uncertainty of the 24-$\mu$m data point,
to account for the highly varied mid-IR spectra of star-forming
galaxies at $z=1.47$.

\subsubsection{Herschel-PACS stacks}

For {\it Herschel}-PACS images from PEP and HerMES surveys, we extract
fluxes using aperture photometry with a radius of 10 and 15\,arcsec at
100 and 160\,$\mu$m, respectively. Aperture photometry is preferred in
PACS mainly due to the uncertainties on the peak of the PSF
(introduced by the telemetry of the {\it Herschel} telescope and by
the asymmetry seen along the scan directions). We have divided the
COSMOS-PEP images by factors of 1.151 and 1.174 at 100 and
160\,$\mu$m, respectively, in order to match the calibration products
used to create the UDS-HerMES image (a change from responsivity FM,5 to FM,6
within the {\it Herschel} Interactive Processing Environment). This
translates into an aperture correction of the order to 30\,per~cent at
those aperture radii. PACS images are in units of Jy\,pixel$^{-1}$, so
we simply use the {\sc idl} routine {\sc aper} to integrate fluxes,
not performing background subtraction as this has been already
removed.

\subsubsection{Herschel-SPIRE stacks}

Following the same recipe used to extract stacked {\it Spitzer}
fluxes, we measure integrated flux densities by simply measuring the
peak in the Gaussian fits. This is valid given that {\it
  Herschel}-SPIRE HerMES images (at 250, 350 and 500\,$\mu$m) are in
Jy\,beam$^{-1}$ units and we are assuming `point-like' stacks. Details
about these images can be found in \citet{Levenson10}.

\subsubsection{AzTEC stacks}

Given that AzTEC images come from different telescopes, the
co-addition of these stacked images needs to include a Gaussian
convolution of the JCMT image ({\sc fwhm} 18\,arcsec;
\citealt{Austermann10}) to match the ASTE resolution (30\,arcsec;
\citealt{Scott08}). Similarly to {\it Herschel}-SPIRE, these maps are
in Jy\,beam$^{-1}$ units, so integrated flux densities are measured by
the peak of a Gaussian fit (same constraints as those used for {\it
  Spitzer} stacks).

\subsubsection{Empirical Calibration}
\label{empirical_cal_section}

To account for possible biases introduced by the way we measure
stacked flux densities, we use the released catalogues from each image
to find the median and scatter (3\,$\sigma$ clipped) of the ratio
between catalogued flux densities (between 5- and 10-$\sigma$) and
fitted Gaussian peaks (aperture photometry for the PACS case).  We
call this ratio $\eta = S_{\rm FIT}/S_{\rm CAT}$.  We note that for
{\it Herschel} and AzTEC images, $\eta$ is within 15\% from unity,
although for {\it Spitzer} images $\eta$ is the value to convert from
ster-radians to beams.  We apply these normalisation factors to the
extracted flux densities in order to make the calibration of each
stacked data point dependent on the released catalogues from each of
the different images (see Table~\ref{table1}). In the cases when we
merge the UDS and COSMOS fields, we use the average correction found
between both fields (usually within 10\,per~cent of each other).
\vspace{1cm}

In summary, for a given number of UDS ($N_{\rm UDS}$) and COSMOS
($N_{\rm COSMOS}$) sources, the calibrated flux density measured from
their co-added stacked signals can be expressed as:

\begin{eqnarray}
  {\rm MED}(S) =& \eta \times
  {\rm MED}
  \left[
    \left(
    S_{i=1,...,N_{\rm UDS}}-B_{\rm MC,UDS}
    \right) 
    + 
    \right. \nonumber \\ & 
    \left. 
    \left(
    S_{j=1,...,N_{\rm COSMOS}}-B_{\rm MC, COSMOS}
    \right)
  \right],
\label{stack_eqn}
\end{eqnarray}

\noindent
where ${\rm MED}$ stands for the median over all the cut-out map
signals $S_i$, and $B_{\rm MC}$ correspond to the Monte-Carlo
simulated sky background obtained by randomising the
positions around the source sample for each field. We find that the
detection significance (fitted peak over local pixel r.m.s.) of each
stacked data point ranges at $<$\,15\,$\sigma$, where the 24-, 250-,
and 350-$\mu$m bands provide the clearest detections.

Errors in our measurements are estimated using the same Monte-Carlo
simulations (100 realisations) to obtain the averaged pixel {\rm rms}
noise for {\it Spitzer}, {\it Herschel}-SPIRE and AzTEC stacks, and
the averaged uncertainty using random aperture photometry on {\it
  Herschel}-PACS stacks. We assume a conservative 10\,per~cent
absolute calibration uncertainty (added in quadrature) for all IR
bands (e.g.\ \citealt{Stansberry06,Swinyard10,Austermann10}). Finally
for consistency, we normalise our estimated uncertainties using the
$\eta$ ratio. 

\subsubsection{Clustering effects}
\label{clus_effec}

The clustering of galaxies can induce a bias on stacking measurements
(\citealt{Bethermin10, Kurczynski10, bBethermin12,
  Heinis13}). \citet{bBethermin12} estimated that for
24\,$\mu$m-selected samples, the level of clustering could increase
the stacked peak flux measurements in the order of 7, 10 and 20 per
cent at 250, 350 and 500\,$\mu$m, respectively. This effect is larger
in lower resolution images and it is seen as wings around the stacked
signals. These wings reflect the effect produced by the excess of
probability to find sources around another one. \citet{bBethermin12}
showed that the shape of this wing is the auto-correlation function
($\omega(\theta)=A_{\rm CF}\theta^{\beta_{\rm CF}}_{\rm CF}$)
convolved by the PSF. The amplitude of this signal thus depends on the
mean flux of the clustered population and the amplitude of
clustering. A preliminary view to the angular correlation function of
HiZELS galaxies at $z=1.47$ ($\sim$\,150--300\,sources\,deg$^{-2}$),
shows it is well-behaved with a power-law ($\beta_{\rm CF}=-0.8$ and
$A_{\rm CF}\approx10-20$ with no clear evidence for a steeper
correlation function at smaller scales) suggesting these galaxies
reside in relatively ‘typical’ dark matter haloes of
$\sim$\,$10^{11}$\,--\,$10^{12}\,$M$_\odot$ (see
e.g.\ \citealt{Sobral10,Geach12}) -- similar (or slightly
  lower) to the ones expected to host 24-$\mu$m sources
  (\citealt{aBethermin12, Wang12}).

In this work, we have neglected the effect of clustering due to three
main reasons: (1) \citet{bBethermin12} estimated clustering effects
using mean stacks while we use median ones. This implies that we
considerably reduce the contribution of objects to clustering
signal, while contribution of sources below the confusion limit stays
the same; (2) H$\alpha$-selected galaxies have stellar masses which
are typically lower than far-IR-selected ones (see later
\S\,\ref{stellar_mass_section_disc}), hence the effect of clustering
is expected to be lower. Actually, we do not see any clear excess
of `wing' emission in our {\it Herschel} stacked images; (3) the
effect that clustering has in stacked SPIRE flux densities using
24\,$\mu$m-selected galaxies is roughly in the range of the global
uncertainties in our work.

%
\begin{figure*}
   \centering
   \includegraphics[scale=0.35]{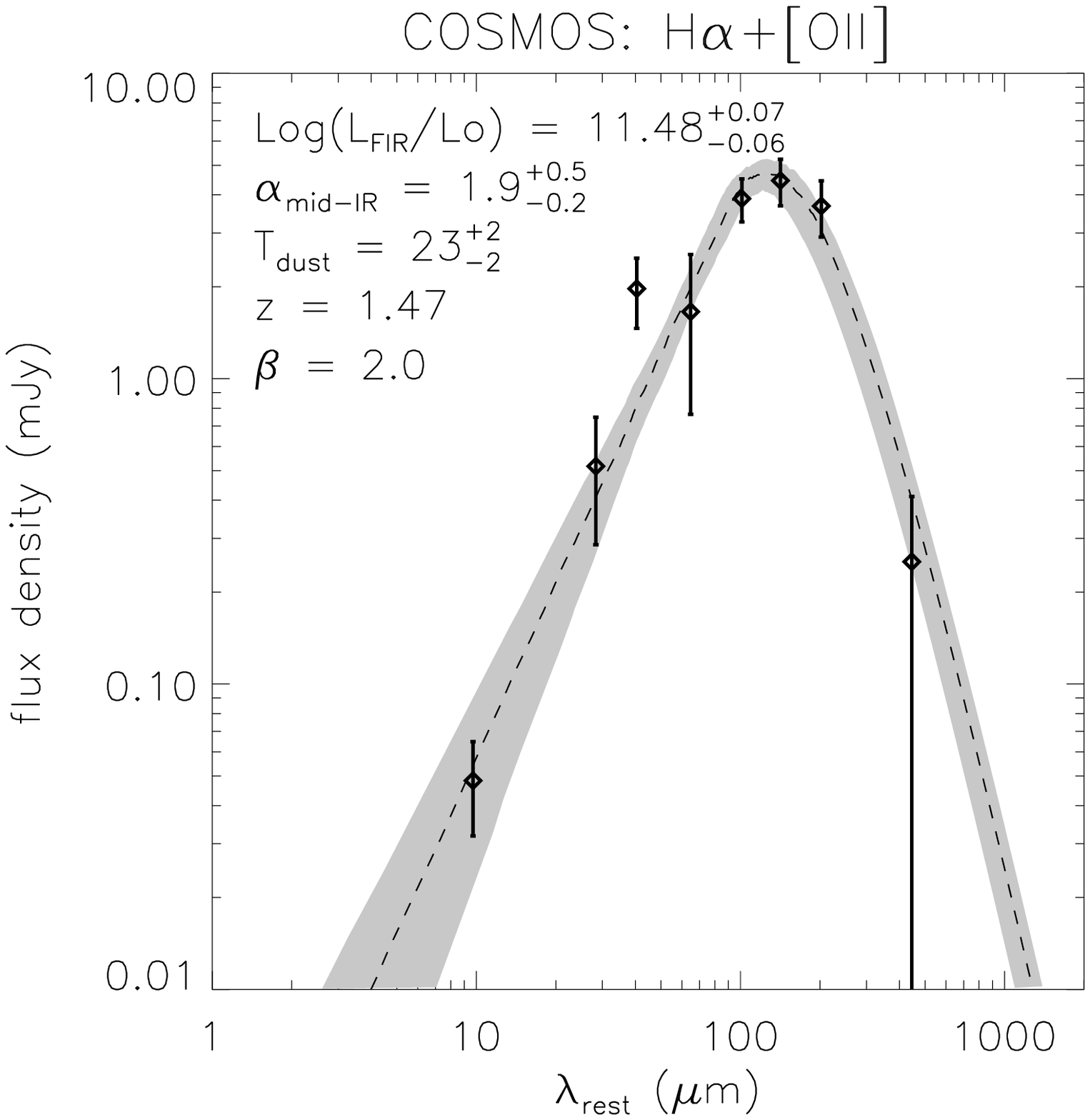}
   \includegraphics[scale=0.35]{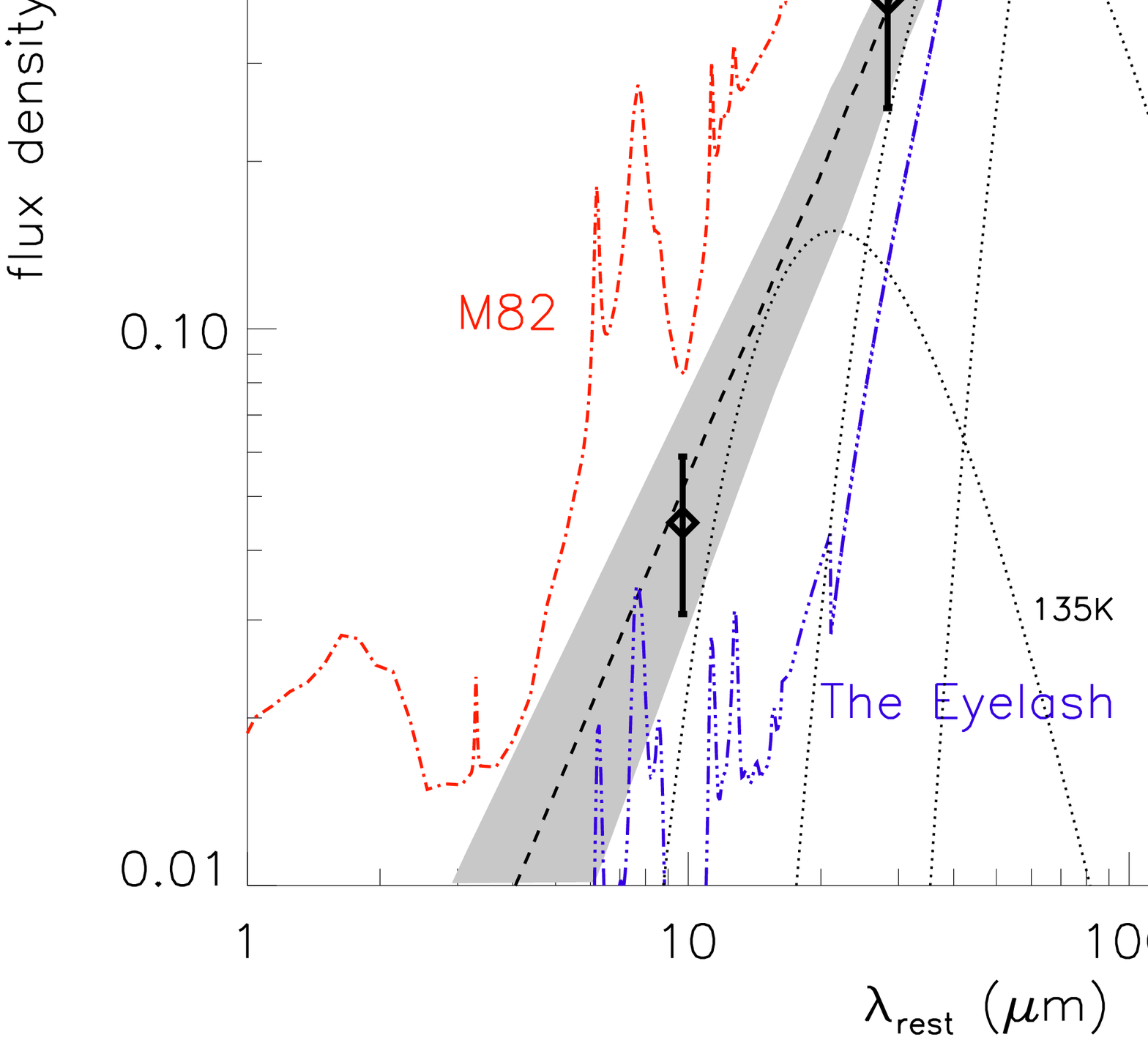}
   \vspace{0.6cm}
   \caption{The stacked far-IR flux densities obtained for different
     HiZELS samples at $z=1.47$. Photometry points come from {\it
       Spitzer} (24 and 70\,$\mu$m), {\it Herschel} (100, 160, 250,
     350 and 500\,$\mu$m) and AzTEC (1100\,$\mu$m) images (see
     Table~\ref{table1} for references). Dashed lines show the
     expected median SED value based on Monte-Carlo realisations
     repeating the SED fits by varying the photometry using the errors
     for each data point (see \S\ref{SED_stack_section}). Derived
     parameters are provided in the inset values (see also
     Table~\ref{table2}) while the plotted shaded areas represent the
     68-per-cent confidence levels. {\bf Left:} the stacked far-IR SED
     for the HiZELS samples with detected ${\rm
       S(H\alpha)>6\times10^{-17}\,erg\,s^{-1}\,cm^{-2}}$ and ${\rm
       S([OII])>1.3\times10^{-17}\,erg\,s^{-1}\,cm^{-2}}$ in the UDS
     ({\it top}) and COSMOS ({\it bottom}) panels. {\bf Right:} the
     large panel shows the stacked SED for the full HiZELS sample,
     compared to two star-forming galaxies: M82 (a local starburst
     galaxy with an SED obtained from a fit to the observed photometry
     presented by \citealt{Silva98}) and the Cosmic Eyelash (a
     gravitationally lensed galaxy at $z=2.3$ -- \citealt{bIvison10}),
     both normalised to arbitrary values. At wavelengths long-ward of
     10\,$\mu$m, the stacked SED can be approximated by the
     superposition of three MBB functions with temperatures of 135, 56
     and 24\,K (using $\beta=2.0$), and luminosity contributions of
     3.5, 16.3 and 80.2\%, respectively.}
   \label{merge_stacks}
\end{figure*}

\subsection{SED fitting}
\label{SED_fitting_section}

We parametrise the stacked SEDs using a modified black-body (MBB)
spectrum in the Rayleigh Jeans regime, but truncated to a power law in
the mid-IR (Wein side). The measured bolometric far-IR luminosities,
$L(8-1000\mu {\rm m})$, are obtained by integrating the SED in
rest-frame frequencies between $\nu_1=0.3$\,THz (1000\,$\mu$m) and
$\nu_2=37.5$\,THz (8\,$\mu$m),

\begin{equation}
  L(8-1000\,\mu{\rm m}) =
  4\pi\,D_{\rm L}^2(z)\,\int_{\nu_{\rm 1}\,/(1+z)}^{\nu_{\rm 2}\,/(1+z)}S_\nu\,d\nu
\end{equation}

\noindent
where the flux density per unit frequency is parametrised as,
$$
  S_\nu(\nu) = 
  A \times \left\{ 
  \begin{array}{ll}
    {\rm MBB}(\nu) & 
    {\rm if}\quad \nu\le\nu_\star      \\
    {\rm MBB}(\nu_\star)\times\nu^{\alpha_{\operatorname{mid-IR}}} & 
    {\rm if}\quad \nu>\nu_\star \\
  \end{array} 
  \right. 
$$

\noindent
and

$$
{\rm MBB}(\nu) = \frac{\nu^{3+\beta}}
{{\rm exp}\left[\frac{h\,\nu}{k\,T_{\rm dust}}(1+z)\right]-1}
$$

Here, $\beta$ is the dust emissivity index
\citep[e.g.][]{Seki80,Dunne11} modulating the Planck function to cover
a wider range of dust temperatures (fixed to $\beta=2$), and
$\nu_\star$ is obtained numerically at

$$
\frac{d\,{\rm Log_{10}(MBB)}}
{d\,{\rm log}_{10}(\nu)}(\nu_\star) =\alpha_{\operatorname{mid-IR}}
$$
\smallskip 

\noindent
to match the slope of the Planck function and the power law at
$\sim$100--200\,$\mu$m. Examples of SED fits applied to our stacked
signals are shown in Fig.~\ref{merge_stacks}.

In our estimates, we assume the redshift is fixed at $z=1.47$
(luminosity distance, $D_{\rm L}(z)=10641$\,Mpc), and parameters $h$
and $k$ refer to Planck and Boltzmann constants, respectively. To fit
the observed stacks we have three free parameters in our model: $A$
(the normalisation of the fits), $T_{\rm dust}$ (as in the MBB
function) and $\alpha_{\operatorname{mid-IR}}$ (the slope in the
mid-IR). The best fit for each parameter is obtained by minimising
$\chi^2$. All quoted errors correspond to the 68-per-cent confidence
levels obtained using an end-to-end Monte-Carlo (100 realisations) fit
to the SED using perturbed photometry based on the estimated errors of
each stack (assuming Gaussianity). 

\subsubsection{Robustness of the SED fit approach}
\label{robustness_of_fits}

Our SED fits parametrise the mid-IR range assuming a simple spectral
slope, which is not physically well-motivated, but at least it does
not bias the derived far-IR luminosities. We demonstrate this by using
stacked signals coming from different parent HiZELS samples, then
fitting them alternating between $\beta$ and
$\alpha_{\operatorname{mid-IR}}$ as fixed and free
parameters. Comparing both outputs, we find that far-IR luminosities
and dust temperatures are not biased by this assumption. The main
reason we preferred to fix $\beta$, rather than
$\alpha_{\operatorname{mid-IR}}$, is the lack of signal-to-noise in
the AzTEC 1100-$\mu$m stacks. This 1100-$\mu$m data point provides the
main constraint on $\beta$, but it is usually only an upper limit. In
our initial analysis we used $\beta=1.5$, but we found that this
produced SED fits which in several cases violated the AzTEC
1100-$\mu$m upper limits, and gave rise to dust temperatures
$\sim$5\,K ($\sim$1$\sigma$) higher than those obtained when fixing
$\alpha_{\operatorname{mid-IR}}=2.0$ with $\beta$ as a free parameter
(see discussion of the $T_{\rm dust}$-$\beta$ relation by
\citealt{Shetty09} and Smith et al.\ {\it in preparation}).

To test the robustness of our SED routine, we used a different fitting
approach involving $\sim$\,7000 SED library models
(\citealt{Siebenmorgen07}) to reproduce the far-IR photometry via a
$\chi^2$ minimisation (\citealt{Symeonidis09,Symeonidis11}).  We find
that our derived far-IR luminosities are systematically 10\,per~cent
lower with respect to this other method. This offset is, however,
within the 1\,$\sigma$ errors. The difference is mostly seen in the
mid-IR part of the spectrum. The SED libraries include prominent PAH
features and a systematic excess of warm dust emission at
$\lambda_{\rm rest}\sim20\,\mu$m, compared to our simple mid-IR power
law. This demonstrates the uncertainties introduced by the SED fits
and the actual capabilities for precisely measuring luminosities in
this work. On the other hand, in terms of dust temperatures we find
that our fits (using $\beta=2$) and the library SED fits are in good
agreement with no clear systematic.

\subsection{The star-formation rate and H$\boldsymbol{\alpha}$ extinction}
\label{section5}

Our study provides a unique opportunity to directly measure the
typical SFR of the HiZELS sample. There is no unbiased SFR indicator
and it is well documented that the use of inconsistent extinction
corrections and SED assumptions are the primary source for the large
scatter seen by different estimators (\citealt{bWijesinghe11}),
especially at high redshifts (e.g.\ \citealt{Hopkins06}). As described
in \S\ref{data_sample}, a key advantage of our HiZELS sample is that
it does not differ significantly from the intrinsic luminosities of
the local star-forming galaxies used to calibrate the SFR indicators
(see Fig.~\ref{sample_figure}). Indeed, the conditions defined by
\citet{Kennicutt09} in \S6.2 of their paper are fully satisfied by the
HiZELS sample. Hence, assuming no cosmic evolution of the parameters
controlling the SFR (e.g.\ the IMF) and the absence of AGN
contamination in the H$\alpha$ luminosities, we can confidently assume
that (\citealt{Kennicutt98}):

\begin{equation}
  {\rm SFR}({\rm M}_\odot\,{\rm yr}^{-1}) = 7.9\times10^{-42}\times
  L({\rm H}\alpha)_{\rm int}
  \quad{\rm erg\,s^{-1}}
\label{eqn2}
\end{equation}

\noindent
where $L({\rm H}\alpha)_{\rm int}$ is the intrinsic H$\alpha$
luminosity (corrected for dust attenuation). This is the canonical
definition for the SFR, assuming solar abundances and a simple
power-law slope for the IMF, $dN/dm\propto m^{-2.35}$
(\citealt{Salpeter55}), integrated between 0.1 and 100\,M$_\odot$. A
way to estimate the extinction for deriving $L({\rm H}\alpha)_{\rm
  int}$ is by combining the measured H$\alpha$ and far-IR luminosities
as follow,

\begin{equation}
  L({\rm H}_\alpha)_{\rm int} = L({\rm H}\alpha)_{\rm obs} +
  a_{\rm FIR} \times L({\rm 8-1000\mu m})\quad {\rm erg\,s^{-1}}
\label{eqn3}
\end{equation}

\noindent
i.e.\ the far-IR luminosity carries the information for the amount of
dust extinction, where $a_{\rm FIR} = (2.5\pm0.6)\times10^{-3}$
(\citealt{Kennicutt09}). Eqn.~\ref{eqn3} can be understood as the
balance of the contributions from unobscured and obscured emission to
the total SFR in a galaxy, and $a_{\rm FIR}$ determines the ratio at
which the components are comparable. The ratio between the two
components can be used to trace the averaged extinction for the
sample,

\begin{equation}
  A_{\rm H\alpha} ({\rm mag}) = 2.5\times {\rm log}_{10}
  \left(
  1+a_{\rm FIR} \frac{L({\rm 8-1000\mu m})}{L({\rm H\alpha})_{\rm obs}}
  \right)
\label{eqn4}
\end{equation}

\noindent
a measure which is expected to be less sensitive to possible ageing
effects given that attenuation decreases with increasing stellar age
(\citealt{Kennicutt09}). As a reference, the typical H$\alpha$
extinction for optically-selected star-forming galaxies in the local
(e.g.\ \citealt{bGarn10}) and high-redshift (e.g.\ \citealt{aGarn10};
\citealt{Sobral12}; \citealt{Stott13}) Universe is $A_{\rm
  H\alpha}\approx 1\,$mag.

Note that since our study uses stacked signals (median properties), we
treat Eqns\,\ref{eqn2}, \ref{eqn3} and \ref{eqn4} in terms of
probability distributions using Monte-Carlo simulations, bootstrapping
the error for the median value of the H$\alpha$ distribution
(Fig.~\ref{sample_figure}) and the measured stacked far-IR flux
errors.

\begin{table*}
  \caption{Different methods to measure the median SFR for the whole
    population of HiZELS galaxies at $z=1.47$ (443 sources). H$\alpha$
    and far-IR luminosities are in erg\,s$^{-1}$ while output values
    are in M$_{\odot}$\,yr$^{-1}$. Our preferred SFR parametrisation
    is in bold at the top of the list, SFR$_{\rm H\alpha,FIR}$ (see
    \S\,\ref{discuss_SFR_ind}; \citealt{Kennicutt09}).  The five
    equations (SFR$_{\rm H\alpha,A_{\rm H\alpha}}$) refer to the
    canonical SFR definition (see Eqn.~\ref{eqn2}) provided by
    \citet{Kennicutt98} using different parametrisation for H$\alpha$
    extinction (scatter of $\sim$\,0.3\,mag) -- taken from
    \citet{bGarn10} and \citet{Sobral12} (see bottom of table). Values
    for $M_\star$ are given in M$_\odot$ (note the 1.21 factor is to
    match our assumed Salpeter IMF and TP-AGB component, see
    \S\,\ref{stellar_mass_section}) and rest-frame $u$--$z$ colour in
    Vega magnitudes. SFR$_{\rm FIR}$ is defined in \citet{Kennicutt98}
    and solely uses the far-IR luminosity to derive the rate of
    star-formation. SFR$_{\rm 24\mu m}$ is defined by \citet{Rieke09}
    in which we have derived $A_{24}=0.54$ and $B_{24}=1.80$ by
    interpolating their Table~1 at $z=1.47$ (see Eqn~14 of their
    paper). Note that ${\rm 4\pi D_L^2\times S_{24\mu m}}$ is in units
    of Jy\,cm$^{2}$, where $D_{\rm L}$ is the luminosity distance and
    $S_{\rm 24\mu m}$ the flux density at 24\,$\mu$m.}  \footnotesize
  \centering
\begin{tabular}{|lll|}
\hline
 & Method & SFR\,(M$_\odot$\,yr$^{-1}$)
\\
\hline
{\bf SFR}$_{\rm\mathbf{H}\boldsymbol{\alpha}\mathbf{,FIR}}$ &
 $\boldsymbol{=}$ 
 {\bf 7.9}
 $\boldsymbol{\times}$ 
 $\boldsymbol{10^{-42}}$ 
 $\boldsymbol{\times}$ 
 $\boldsymbol{[L({\rm H}\alpha)_{\rm obs}}$ 
 $\boldsymbol{+}$ 
 $\boldsymbol{a_{\rm FIR}}$ 
 $\boldsymbol{\times}$ 
 $\boldsymbol{L(8$ $-$ $1000\mu {\rm m})]}$ &
 $\boldsymbol{=32 \pm 5}$ \\
SFR$_{\rm H\alpha,A_{H\alpha}=0}$&
= 7.9 $\times$ $10^{-42}$ $\times$ $L({\rm H}\alpha)_{\rm obs}$ &
$=13.1 \pm 0.3$ \\
SFR$_{\rm H\alpha,A_{H\alpha}=1}$ &
= $7.9$ $\times$ $10^{-42}$ $\times$ $ L({\rm H}\alpha)_{\rm obs}$ $\times$ $10^{0.4}$ &
$=31.9 \pm 0.8$ \\
SFR$_{\rm H\alpha,A_{H\alpha}([OII]/H\alpha)}$ &
= $7.9$ $\times$ $10^{-42}$ $\times$ $ L({\rm H}\alpha)_{\rm obs}$ $\times$ $10^{0.4\,A_{\rm H\alpha}(\rm [OII]/H\alpha)}$ &
$=28.9 \pm 1.4$\\
SFR$_{\rm H\alpha,A_{H\alpha}(M_\star)}$ &
= $7.9$ $\times$ $10^{-42}$ $\times$ $ L({\rm H}\alpha)_{\rm obs}$ $\times$ $10^{0.4\,A_{\rm H\alpha}(\rm M_\star)}$ &
$=24.8 \pm 0.7$ \\
SFR$_{\rm H\alpha,A_{H\alpha}({\rm [u-z]_{rest}})}$ &
= $7.9$ $\times$ $10^{-42}$ $\times$ $ L({\rm H}\alpha)_{\rm obs}$ $\times$ $10^{0.4\,A_{\rm H\alpha}({\rm [u-z]_{rest}})}$ &
$=22.9 \pm 0.6$ \\
SFR$_{\rm FIR}$ &
= $4.5$ $\times$ $10^{-44}$ $\times$ $\,L(8$ $-$ $1000\mu{\rm m}) $ &
$=44^{+ 4}_{- 6}$\\
SFR$_{\rm 24\mu m}$ &
= ${\rm 10^{A_{24}+B_{24}\times[Log(4\pi D_L^2\times S_{24\mu m})-53]}}$ &
$=90 \pm 18$ \\
\hline
where & &
\\
$A_{\rm H\alpha}(\rm [OII]/H\alpha)$ &
=  $-4.30\,X^4 -11.30\,X^3 -7.39\,X^2 -2.94\,X + 0.31$\quad 
&{\rm using}\quad$X={\rm Log_{10}([OII]/H\alpha)}$ 
\\
$A_{\rm H\alpha}(\rm M_\star)$ &
=  $-0.09\,X^3 + 0.11\,X^2 + 0.77\,X + 0.91$\quad 
&{\rm using}\quad$X={\rm Log}_{10}(\rm M_\star/10^{10}/1.21)$
\\
$A_{\rm H\alpha}({\rm [u-z]_{rest}})$ &
=  $-0.092\,X^3 + 0.671\,X^2 -0.952\,X + 0.875$\quad 
&{\rm using}\quad$X=(u-z)_{\rm rest}$
\\
\hline
\end{tabular}
\label{sfrs_all_sample}
\end{table*}

\subsection{The global far-IR properties of the star-forming HiZELS population}
\label{compare_stacks}

In the left hand panel of Fig.~\ref{merge_stacks}, we compare the
stacked far-IR fluxes of all those UDS and COSMOS HiZELS galaxies with
narrow-band H$\alpha$ and \OII\ detections (see
Fig.~\ref{sample_figure}).  Both samples produce consistent far-IR
fluxes to within the uncertainties, suggesting that we can merge them
to increase the overall sample size and to allow an alternative sample
selection to investigate the far-IR properties as a function of
different observed physical parameters. The stacked fluxes obtained
from the merged full sample are shown in Fig.~\ref{merge_stacks} and
derived parameters presented in Table~\ref{table5}\,\&\,\ref{table2}.

The observed SEDs peak roughly at 280\,$\mu$m (rest-frame\,
$\sim$113\,$\mu$m) corresponding to a dust temperature, $T_{\rm
  dust}\,\sim$\,24\,K. As noted in \S\,\ref{robustness_of_fits}, this
value increases to $\sim$\,30\,K when $\beta=1.5$ is used. The
assumption of a simple mid-IR power law is useful to mitigate the
larger uncertainties at shorter wavelengths (especially on {\it
  Spitzer}-70\,$\mu$m and PACS-100/160\,$\mu$m photometry), where we
usually find $\alpha_{\operatorname{mid-IR}}=2$, e.g.\ similar to that
observed in M82. We find that the use of a modified black-body is
essential to fit the stacked SED. This makes perfect sense since the
stacked fluxes include the modified black-body emission from each
independent galaxy as well as the broadening introduced by stacking
H$\alpha$ emitters with different dust temperatures (the expected
$z=1.47\pm0.02$ distribution has a negligible effect compared to these
two effects). In our SED fits, this broadening is basically modulated
by $\beta$ and $\alpha_{\operatorname{mid-IR}}$.  Alternatively, in
Fig.~\ref{merge_stacks} we show that the typical HiZELS SED (at
$\lambda>10\,\mu$m) can be approximated by the composition of three
MBB functions with $T_{\rm dust} = 24$, 56, and 135\,K (using
$\beta=2.0$), contributing in 80.2, 16.3 and 3.5\,\% to the total
far-IR luminosity.


The derived median far-IR luminosity for the whole merged sample is
$L({\rm 8-1000\mu m})= 10^{11.41_{-0.06}^{+0.04}}$\,L$_{\odot}$,
i.e.\ our H$\alpha$ emitters are typically luminous infrared galaxies
(LIRGs) at $z=1.47$. The ratio between the far-IR and the observed
H$\alpha$ luminosities is $\sim$1000:2, similar to the $a_{\rm FIR}$
factor from Eqn~\ref{eqn3}. This implies that these two components
(unobscured and obscured) have comparable contributions to the total
SFR. Using Eqn.~\ref{eqn4}, we derive a median $A_{\rm
  H\alpha}=1.0\pm0.2$\,mag, in good agreement with typical values seen in
local and high-$z$ star-forming galaxies (\citealt{bGarn10,Sobral12,
  Stott13}).

Given that our sample has been primarily selected by its H$\alpha$
power, we cannot assume HiZELS galaxies work as calorimeters
(\citealt{Lacki10}). Indeed, this is the main reason we prefer the use
of a combination of H$\alpha$ and far-IR luminosities to derive total
SFRs (\citealt{Kennicutt09}), rather than H$\alpha$ or far-IR
luminosities alone (\citealt{Kennicutt98}).  In
Table~\ref{sfrs_all_sample}, we present the different methods used to
compare SFR estimates in this work.  For example, assuming a simple
$A_{\rm H\alpha}=1$ to get the intrinsic H$\alpha$ luminosity, the
SFR$_{\rm H\alpha,A_{H\alpha}=1}$ is roughly within 1.4--3$\times$ of
that derived using SFR$_{\rm FIR}$ or SFR$_{\rm 24\mu m}$ stacks.
This simple comparison reinforces the fact that a non-negligible
fraction of the starlight has escaped from these galaxies. We find
that the SFR$_{\rm FIR}$ are typically 1.5$\times$ larger than those
expected from SFR$_{\rm H\alpha,FIR}$ (\S\ref{discuss_SFR_ind}). In
particular, we note that the use of the 24-$\mu$m flux density as a
SFR indicator at $z=1.47$ is relatively uncertain compared to
SFR$_{\rm FIR}$. This is due to the combination of being estimated
using a single photometry point and due to the large uncertainty
induced by the Silicate absorption band at 9.8\,$\mu$m, Polycyclic
Aromatic Hydrocarbon (PAH) line emissions and possible AGN power law
components redshifted into the 24\,$\mu$m band.

%
\begin{figure}
   \centering
   \includegraphics[scale=0.45]{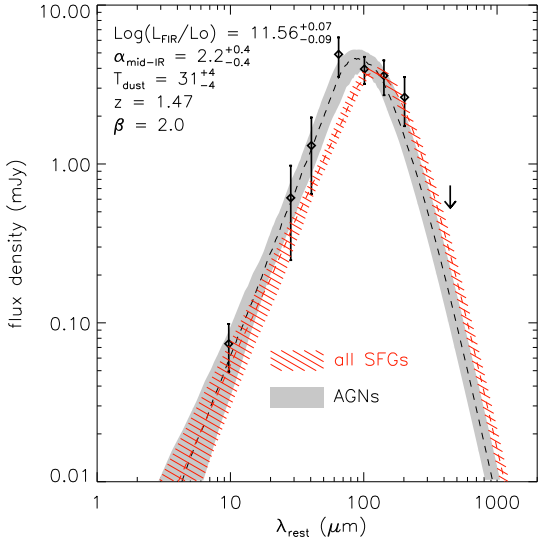}
   \caption{Stacked SED for all possible AGN (grey shaded region),
     including radio-loud, X-ray-detected and mid-IR identified ones
     (see \S\ref{AGN_removal_section} for details). The derived far-IR
     properties for the AGN sample are inset at the top-left.  The
     stacked AGN SED is compared to the one obtained from all HiZELS
     star-forming galaxies (red line-filled; presented in
     Fig.~\ref{merge_stacks}-{\it right}).}
   \label{AGN_sed}
\end{figure}

\subsection{AGN contamination}
\label{sect_AGN_cont}

In \S\ref{AGN_removal_section} we explained the conservative method we
have used to clean the HiZELS sample of possible AGN
contamination. Note that this method might have classified some
powerful star-forming galaxies as AGN. Using our X-ray/mid-IR/radio
criteria we have identified a total of 70 possible AGN within our
HiZELS sample.

Using the same stacking approach explained above, we show in
Fig.~\ref{AGN_sed} the typical far-IR SED for the HiZELS galaxies
classified as AGN. We compare the full stacked star-forming sample
(presented in Fig.~\ref{merge_stacks}-{\it right}) with respect to
that of the AGN population, finding that the typical far-IR
luminosities for AGN are slightly larger
($\sim$\,$10^{11.56\pm0.08}\,$L$_\odot$). There is evidence for warmer
dust temperatures ($\Delta T_{\rm dust}\sim$\,7\,K) with respect to
the star-forming galaxies. It is interesting to see AGN do not
dominate the mid-IR part of the stacked SEDs -- even though we would
expect some mid-IR emission coming from the central torus-like region
surrounding the AGN, introducing an excess at 24\,$\mu$m (9.7\,$\mu$m,
rest-frame).

As a sanity check, we performed the same analysis, leaving the
identified AGN in the sample. We find that there is no significant
variation of the results presented in
Figs.~\ref{stack_properties_fig1} and \ref{stack_properties_fig2}, and
all general tracks are maintained within a fraction of the 1-$\sigma$
errors.

We conclude that given the number of identified AGN is small with
respect to the whole sample ($\sim$\,15\,per~cent), possible biases
introduced in median stacked signals are minimised unless AGN are the
dominant population. There is no evidence suggesting such scenario.

\subsection{Dependency of SFR on H$\mathbf{\alpha}$ luminosity and
  stellar mass}
\label{stellar_mass_section}

\begin{figure*}
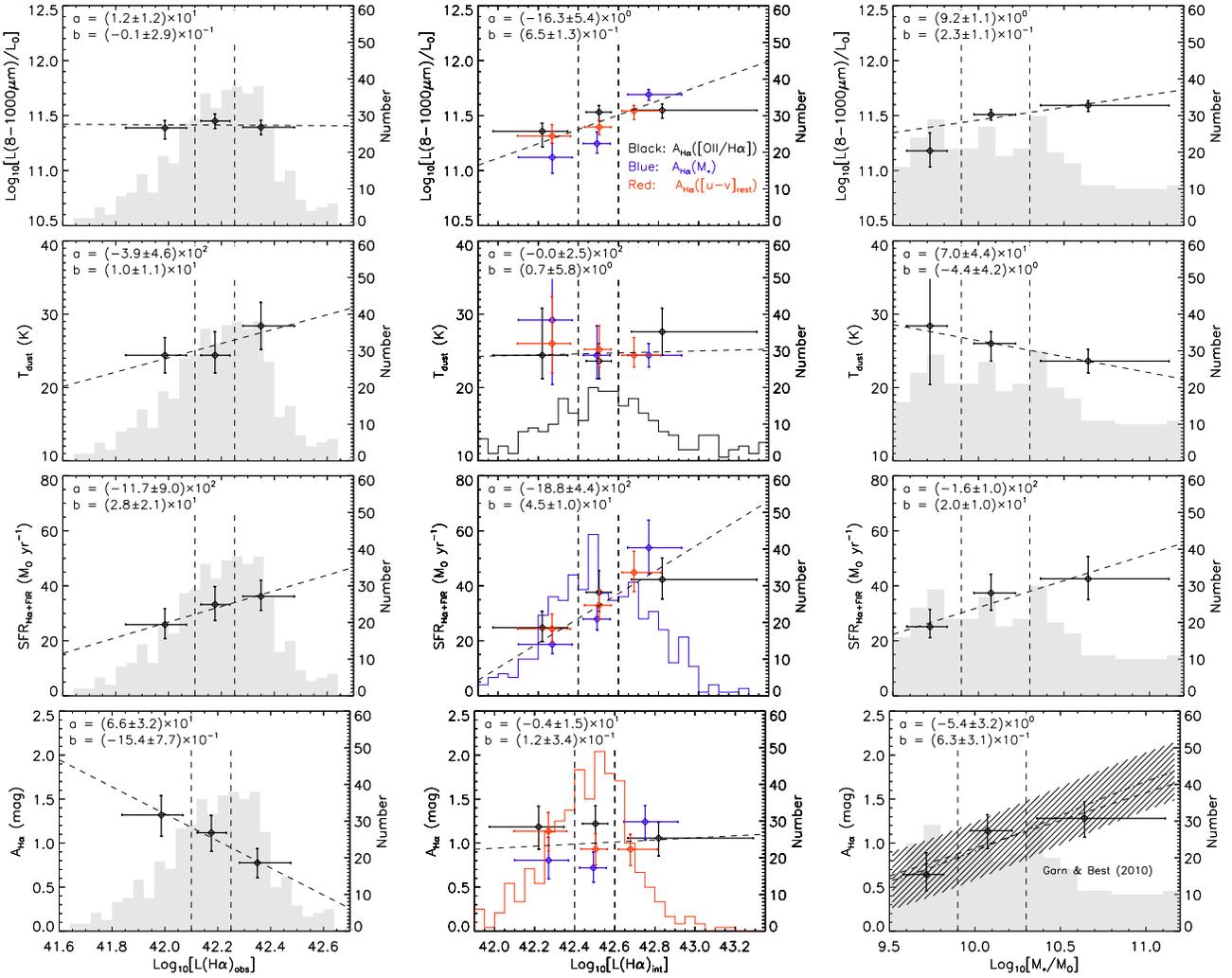

   \centering
   \includegraphics[scale=0.41]{FIN_plot_hist_SFR_Obs_LHa.ps0}
   \includegraphics[scale=0.41]{FIN_plot_hist_SFR_Int_LHa_ALL.ps0}
   \includegraphics[scale=0.41]{FIN_plot_hist_SFR_Mstar.ps0}\\
   \includegraphics[scale=0.41]{FIN_plot_hist_SFR_Obs_LHa.ps1}
   \includegraphics[scale=0.41]{FIN_plot_hist_SFR_Int_LHa_ALL.ps1}
   \includegraphics[scale=0.41]{FIN_plot_hist_SFR_Mstar.ps1}\\
   \includegraphics[scale=0.41]{FIN_plot_hist_SFR_Obs_LHa.ps2}
   \includegraphics[scale=0.41]{FIN_plot_hist_SFR_Int_LHa_ALL.ps2}
   \includegraphics[scale=0.41]{FIN_plot_hist_SFR_Mstar.ps2}\\
   \includegraphics[scale=0.41]{FIN_plot_hist_SFR_Obs_LHa.ps3}
   \includegraphics[scale=0.41]{FIN_plot_hist_SFR_Int_LHa_ALL.ps3}
   \includegraphics[scale=0.41]{FIN_plot_hist_SFR_Mstar.ps3}
   \vspace{0.6cm}
   \caption{Plots showing (from top to bottom) the bolometric far-IR
     emission (8--1000\,$\mu$m), the fitted dust temperature, the
     SFR$_{\rm H\alpha,FIR}$ (see Table~\ref{sfrs_all_sample}) and the
     measured H$\alpha$ extinction using Eqn.~\ref{eqn4} as a function
     of (from left to right) observed L(H$\alpha$), intrinsic
     L(H$\alpha$) (using the recipes provided in
     Table~\ref{sfrs_all_sample}) and stellar mass (see
     \S~\ref{stellar_mass_section}). The histograms show the parameter
     distribution for the stacked population divided in 3 bins as
     defined by the vertical dashed lines. The coloured histograms in
     the middle panels show the ${\rm L(H\alpha)_{int}}$ distributions
     using three different extinction corrections (following the same
     colours as the legend in the top-middle panel). The {\it
       bottom-right} panel includes the $A_{\rm H\alpha}({\rm
       M}_\star)$ parametrisation described in \citet{bGarn10} (see
     Table~\ref{sfrs_all_sample}). Derived quantities and errors
     (1-$\sigma$ enclosing 68\%) are based on an end-to-end
     Monte-Carlo realisation of the fitting SED routine using the
     uncertainties measured in each of the stacked signals
     (\S\ref{SED_fitting_section}), including the uncertainty in
     $a_{\rm FIR}$ from Eqn.~\ref{eqn3}. These data
       points are presented in Table~\ref{table2}. Dotted lines show
       simple linear fits to the stacked data points, $y(x)=a+b\times
       x$, where best fit parameters are inset at the top left of each
       figure and errors are 1-$\sigma$ uncertainty estimates for the
       returned parameters (obtained using {\sc linfit} routine in
       {\sc idl}).}
   \label{stack_properties_fig1}
\end{figure*}

\begin{figure*}
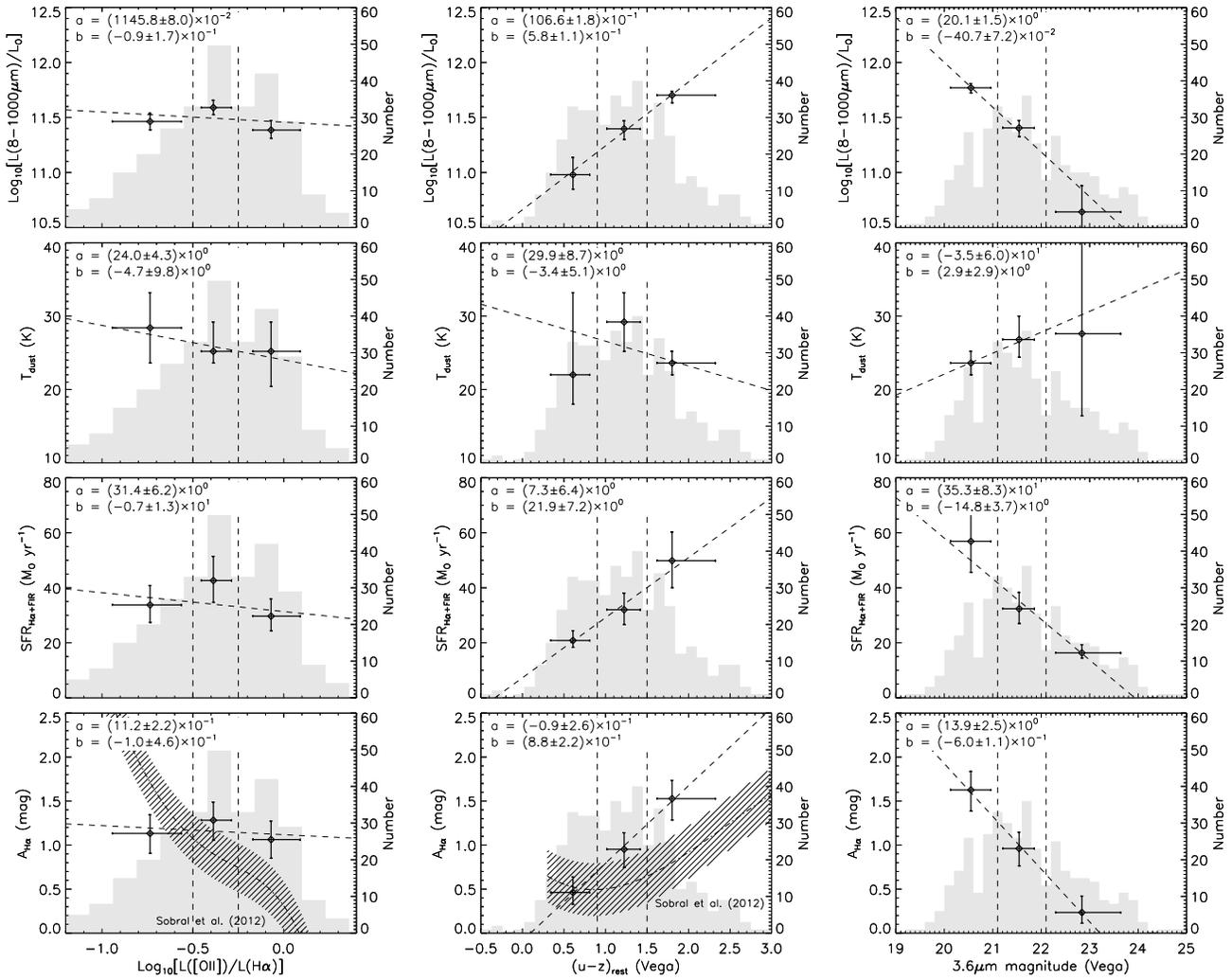

   \centering
   \includegraphics[scale=0.41]{FIN_plot_hist_SFR_LOG_OII_LHa_ratio.ps0}
   \includegraphics[scale=0.41]{FIN_plot_hist_SFR_u-z_rest.ps0}
   \includegraphics[scale=0.41]{FIN_plot_hist_SFR_3.6um.ps0} \\
   \includegraphics[scale=0.41]{FIN_plot_hist_SFR_LOG_OII_LHa_ratio.ps1}
   \includegraphics[scale=0.41]{FIN_plot_hist_SFR_u-z_rest.ps1}
   \includegraphics[scale=0.41]{FIN_plot_hist_SFR_3.6um.ps1} \\
   \includegraphics[scale=0.41]{FIN_plot_hist_SFR_LOG_OII_LHa_ratio.ps2}
   \includegraphics[scale=0.41]{FIN_plot_hist_SFR_u-z_rest.ps2}
   \includegraphics[scale=0.41]{FIN_plot_hist_SFR_3.6um.ps2} \\
   \includegraphics[scale=0.41]{FIN_plot_hist_SFR_LOG_OII_LHa_ratio.ps3}
   \includegraphics[scale=0.41]{FIN_plot_hist_SFR_u-z_rest.ps3}
   \includegraphics[scale=0.41]{FIN_plot_hist_SFR_3.6um.ps3} \\
   \vspace{0.6cm}
   \caption{As Fig.~\ref{stack_properties_fig1}, but this time showing
     the derived properties as a function of the ${\rm
       log}_{10}[L($\OII$)/L({\rm H\alpha})]$ luminosity ratio,
     rest-frame ${[u-z]_{\rm rest}}$ colour, and observed {\it
       Spitzer}-3.6-$\mu$m photometry.  On the bottom figures we
     compare our results with the $A_{\rm H\alpha}$ parametrisations
     presented by \citet{Sobral12}. We conclude that
       $L($\OII$)/L({\rm H\alpha})$ does not correlate with the
       H$\alpha$ extinction (defined in Eqn.~\ref{eqn3}) as predicted
       by \citet{Sobral12} using Balmer decrements. Significant
       correlations are found in terms of the simple observables,
       ${(u-z)_{\rm rest}}$ and 3.6-$\mu$m, which can be used to
       parametrise the obscured star-formation in these H$\alpha$
       emitters (see Eqn.\ref{eqn_sfr_mass3}). In particular, the
       bottom-middle panel suggests the appearance of a highly
       extinguished component in red galaxies which is not totally
       evidenced by optical studies (e.g.\ \citealt{Sobral12}).}
   \label{stack_properties_fig2}
\end{figure*}

In Fig.~\ref{stack_properties_fig1}, we explore far-IR-derived
quantities in order to understand the mechanisms controlling the
star-formation activity of HiZELS at $z=1.47$. We present how the
far-IR luminosity $L(\rm 8-1000\,\mu m)$, the dust temperature, the
derived SFR$_{\rm H\alpha,FIR}$ (see Table~\ref{sfrs_all_sample}) and
the H$\alpha$ extinction (Eqn.~\ref{eqn4}) correlate with the observed
(and intrinsic) H$\alpha$ luminosity and stellar mass. We are able to
create three bins for each parameter, with sufficiently large and
similar number of sources in each bin to define the SED accurately. We
find a relatively mild dependency for far-IR luminosity on observed
$L({\rm H\alpha})$, and stellar mass, but a significant linear slope
(at 5-$\sigma$ significance) on intrinsic $L({\rm H\alpha})$. The
SED-fitting uncertainties hide any trends in dust temperature, $T_{\rm
  dust}$, but values are in rough agreement with previous {\it
  Herschel}-selected samples which are assumed to be mostly dominated
by obscured star formation \citep[e.g.][]{Hwang10,
  Amblard10}. SFR$_{\rm H\alpha,FIR}$ increases as a function of all
three parameters, with $L({\rm H\alpha})_{\rm int}$ showing the
strongest dependence (a 4.5-$\sigma$ significance in linear slope). We
find an anti-correlation between H$\alpha$ extinction and H$\alpha$
luminosity, expected given the form of Eqn.\,\ref{eqn4}. Extinction
increases at large stellar masses in agreement with the trend seen by
\citet{bGarn10} (see Fig.~\ref{stack_properties_fig1}-{\it
  bottom-right} panel). As with $T_{\rm dust}$, we do not find any
strong correlation for $\alpha_{\operatorname{mid-IR}}$. All the
stacked fluxes and derived parameters are presented in
Table~\ref{table5}\,\&\,\ref{table2}.

%
\begin{figure}
   \centering
   \includegraphics[scale=0.48]{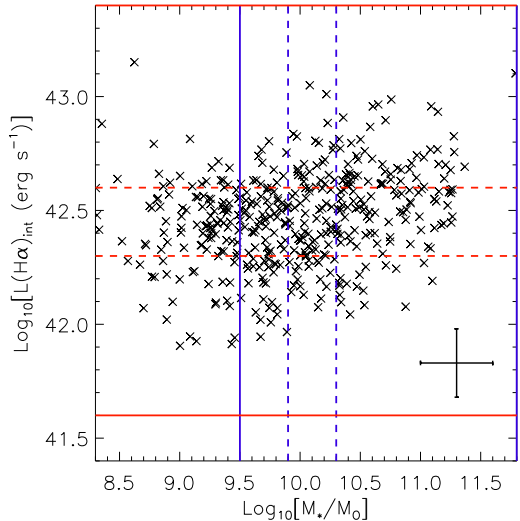}
   \caption{The distribution of intrinsic H$\alpha$ luminosities
     versus SED-fitted stellar masses (see
     \S\,\ref{stellar_mass_section}).  $L({\rm H}\alpha)_{\rm int}$
     are obtained using the recipe provided by \citet{Sobral12} which
     parametrise $A_{\rm H\alpha}$ as a function of the observed
     rest-frame $u$-$z$ colour (see Table~\ref{sfrs_all_sample}).  The
     typical error is shown at the bottom right of the figure. Solid
     blue lines define the minimum/maximum reliable range used for
     stacking. Dashed blue lines define the bins presented in
     Fig.~\ref{stack_properties_fig1}.}
   \label{LHa_mass_diagram}
\end{figure}

%

%

\subsubsection{Stellar masses}

Stellar masses are derived via SED-fitting routines using the method
described by \citet{Sobral11}, making use of the UV to 
  {\it Spitzer}-IRAC near-IR photometry available in the COSMOS and
UDS fields (see Fig.~\ref{LHa_mass_diagram}). We have
included the TP-AGB contribution (\citealt{Bruzual07} templates) as it
has a significant effect for young stellar populations at $z\approx
1.5$, compared to masses derived with previous \citet{Bruzual03} SED
libraries ($\sim$\,1.5$\times$ lower values). To be consistent, the
previously published HiZELS stellar masses (\citealt{Sobral11}) are
divided by a factor of 0.55 to change from a \citet{Chabrier03} IMF to
a Salpeter IMF. Based on the sensitivity of the broadband photometry,
we find that we are unable to confidently constrain stellar masses
lower than $10^{9.5}\,{\rm M}_\odot$, which defines our lower mass
limit for this particular analysis. By excluding sources with
M$_\star<10^{9.5}\,{\rm M}_\odot$ and poor $\chi^2$ fits to the
optical/near-IR SEDs, we remove $\sim$\,25\,per cent of the full
sample. Note that this cut in the sample is only performed to explore
correlations as a function of stellar mass, not in the other
analyses. The low stellar mass population tend to be biased towards
the highest star-formation efficiencies (see later Fig.\ref{sSFR_fig})
-- a population which might not be the most representative one.

The distribution of intrinsic H$\alpha$ luminosities (using $A_{\rm
  H\alpha}([{\rm u-z}]_{\rm rest})$; see different parametrisations in
Table~\ref{sfrs_all_sample}) versus stellar masses can be seen in
Fig.~\ref{LHa_mass_diagram}.  We see a large scatter
  between these two parameters (a Pearson's product moment correlation
  of 0.23 giving a 5\% probability of correlation), probably due to
  the relatively narrow range of luminosities and the large
  uncertainties associated to the H$\alpha$ extinction corrections
  ($\Delta A_{\rm H\alpha}\approx0.3$\,mag in scatter;
  \citealt{Sobral12}).

\subsection{Exploring other dependencies}

Given the difficulty of detecting H$\beta$ at high-$z$
(e.g.\ \citealt{Stott13}), various parametrisations were created by
\citet{Sobral12} -- based on SDSS data -- to describe the H$\alpha$
extinction at $z\sim 1.5$. In this section, we explore the
parametrisation of $A_{\rm H\alpha}$ as a function of the
\OII/H$\alpha$ ratio, rest-frame $u-z$ colour and observed 3.6\,$\mu$m
magnitudes (see Fig.~\ref{stack_properties_fig2}). By looking at the
significance of the linear fit's slopes shown in
Fig.~\ref{stack_properties_fig2}, we find that the far-IR luminosity
is not traced by the \OII/H$\alpha$ ratio, but depends strongly on
intrinsic rest-frame $(u-z)$ colour and observed 3.6\,$\mu$m
photometry (rest-frame $\sim$\,K-band).

\OII\ is a collisionally excited doublet ($\lambda\lambda$372.6,
372.9\,nm) used as a SFR tracer for intermediate-redshift star-forming
galaxies (e.g.\ \citealt{Hayashi13}). It is sensitive to the abundance
and the ionisation state of the gas, and its luminosity is less
directly coupled to the radiation fields from H\,{\sc ii} regions than
H$\alpha$. Empirical evidence has shown a typical observed ratio of
$L($\OII$)/L({\rm H\alpha})=0.23$ in massive galaxies
\citep{Hopkins03}. \citet{Sobral12} proposed this ratio could be used
to parametrise the Balmer decrement (i.e.\ as a tracer of extinction)
after finding they correlate (removing metallicity dependencies) in a
carefully selected SDSS galaxy sample. In
Fig.~\ref{stack_properties_fig2}, we show that this ratio does not
have any clear correlation with our far-IR-derived parameters. This
result might be explained by the fact that the $A_{\rm
  H\alpha}($\OII$/{\rm H\alpha})$ parametrisation induces a
significantly broader extinction-corrected ${\rm
  Log_{10}[L(H\alpha)_{int}]}$ histogram distribution (shown in the
middle panels of Fig.~\ref{stack_properties_fig1}). We suggest that
this optical line ratio is a poor tracer of the far-IR luminosity
and/or H$\alpha$ extinction, but only sensitive to a relatively small
number of unobscured star-forming regions, missing a significant
fraction of the obscured star-forming clouds.

We have also looked at the correlation between $A_{\rm H\alpha}$ and
galaxy colour (e.g.\ \citealt{Sobral12}), finding a relatively strong
dependence in rest-frame $(u-z)_{\rm rest}$ colour (linear slope at
4-$\sigma$ significance). This correlation suggests that red colours
in a sample of young star-forming galaxies (H$\alpha$-detected) are
evidence of systems suffering large extinctions. We find our trend
agrees with \citeauthor{Sobral12}'s previous $A_{\rm
  H\alpha}([u-z]_{\rm rest})$ parametrisation, although we see a
higher level of extinction in redder colours. This suggest the
appearance of a heavily obscured component not accounted by the
previous optical measurements but revealed in the far-IR.

Finally, we explored how the far-IR luminosity depends on observed
{\it Spitzer}-3.6\,$\mu$m magnitudes. The observed 3.6\,$\mu$m
corresponds roughly to the rest-frame $K$-band at $z=1.47$, hence a
proxy for the stellar mass of the galaxy. The $K$-band versus $A_{\rm
  H\alpha}$ correlation is found to be slightly stronger than on
stellar mass, suggesting this {\it Spitzer} band traces the old
stellar populations but also some of the recent `young' star-forming
population.

\section{Discussion}
\label{discussion_section}

\subsection{SFR indicators}
\label{discuss_SFR_ind}

In this work, it is of major importance to use the best SFR estimator
available in the literature. We note that HiZELS galaxies were
selected by their H$\alpha$ emission (in the rest-frame $R$ band), so
by definition considerable starlight must have escaped from these
star-forming galaxies. This implies HiZELS galaxies do not work as calorimeters
(\citealt{Lacki10}). Indeed, we have shown that SFR$_{\rm H\alpha,
  A_{\rm H\alpha}=0}$ (the H$\alpha$-derived SFR not corrected by
extinction) is roughly within a factor of 3 to 4 to the far-IR-derived
SFRs, clearly pointing out the importance of using a combination of
H$\alpha$ and far-IR measurements to obtain the total SFR (see also
\citealt{Iglesias-Paramo06, bWijesinghe11}).

We find that our sample has a typical $L({\rm H\alpha})_{\rm
  obs}/L(8-1000\,\mu{\rm m})\approx 2/1000$ ratio (similar to $a_{\rm
  FIR}=2.5\times10^{-3}$, see Eqn.~\ref{eqn3}). Considering also the
fact that HiZELS H$\alpha$ luminosities are similar (although at
$z=1.47$) to those used by \citet{Kennicutt09} to determine the local
SFR calibrations, we propose that the combination of H$\alpha$ and
far-IR luminosities (SFR$_{\rm H\alpha,FIR}$; \citealt{Kennicutt09})
provides the most reliable SFR estimate for our particular study.

The HiZELS population at $z=1.47$ is composed in its majority of LIRGs
($10^{11}<{\rm L(8-1000\mu m)/L_\odot}<10^{12}$) with SFR$_{\rm
  H\alpha,FIR}=32\pm5$\,M$_\odot$\,yr$^{-1}$. They present typical
H$\alpha$ extinctions ($A_{\rm H\alpha}=1.0\pm0.2$\,mag) which are
similar to those seen in local star-forming galaxies (i.e.\ in
galaxies with lower SFRs). These results demonstrate the little
evolution of the global H$\alpha$ extinction properties previously
seen by \citet{Sobral12} and \citet{Stott13} in HiZELS galaxies at
$z=1.47$.

%
\begin{figure}
   \centering
   \includegraphics[scale=0.55]{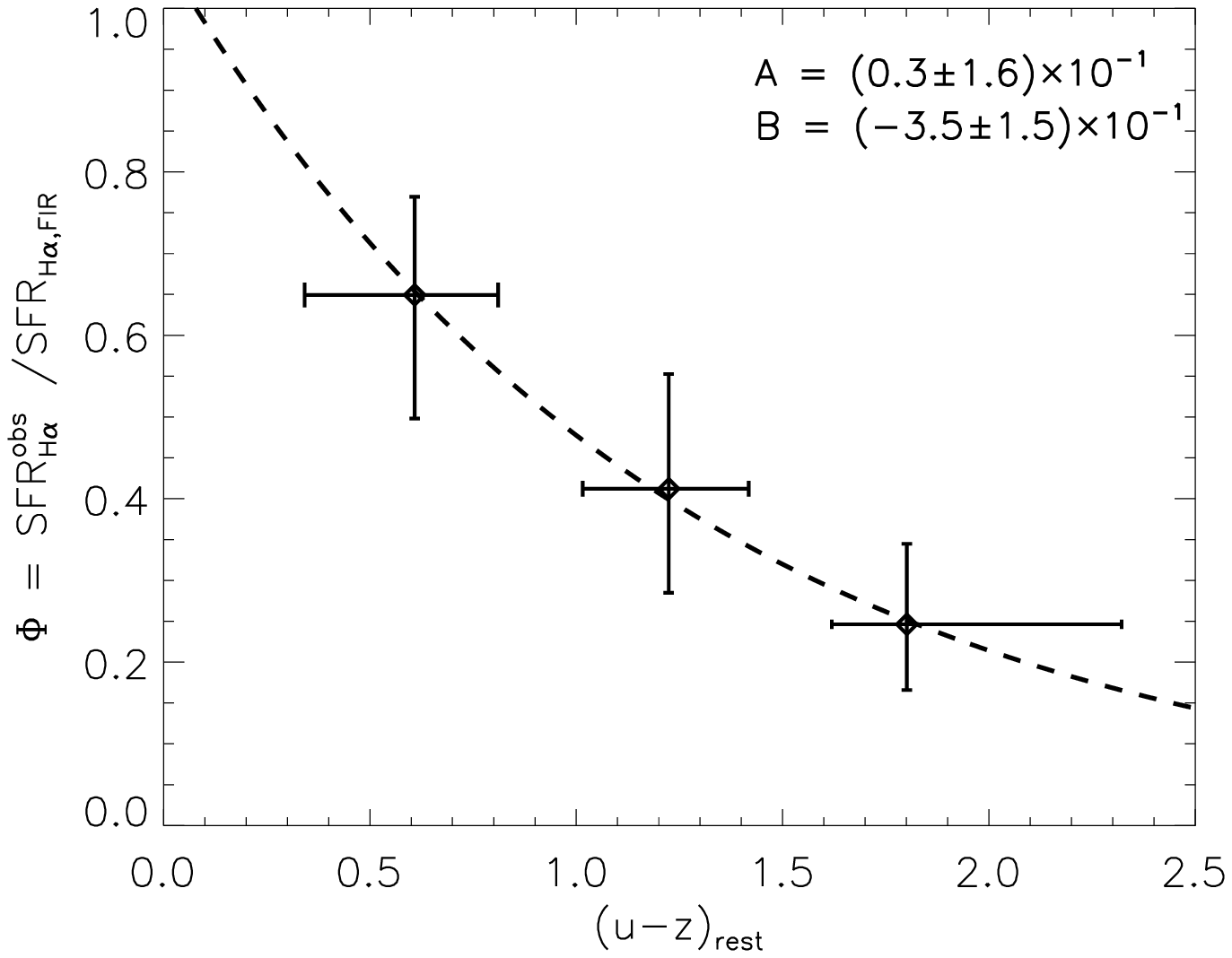}
   \includegraphics[scale=0.55]{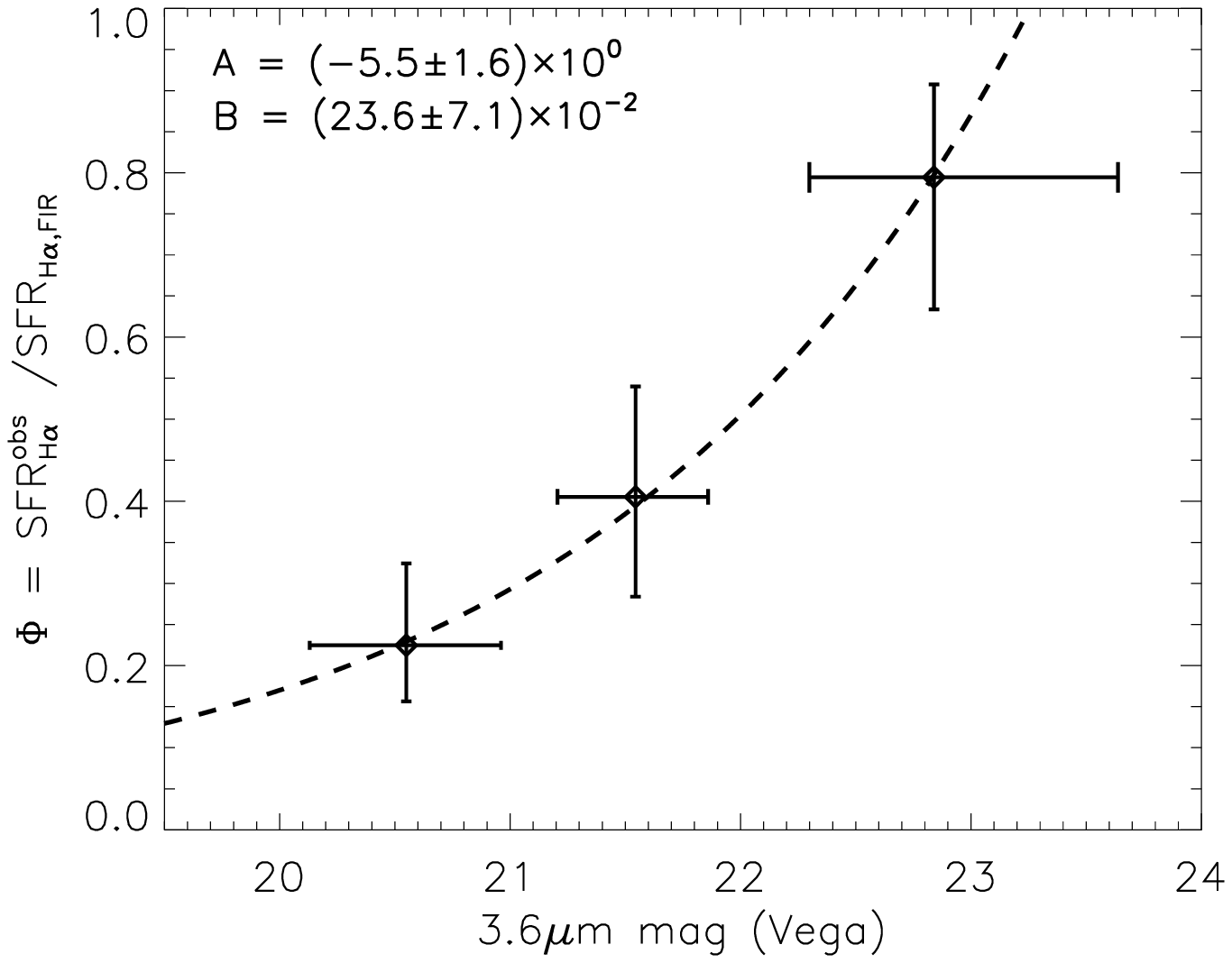}
   \includegraphics[scale=0.55]{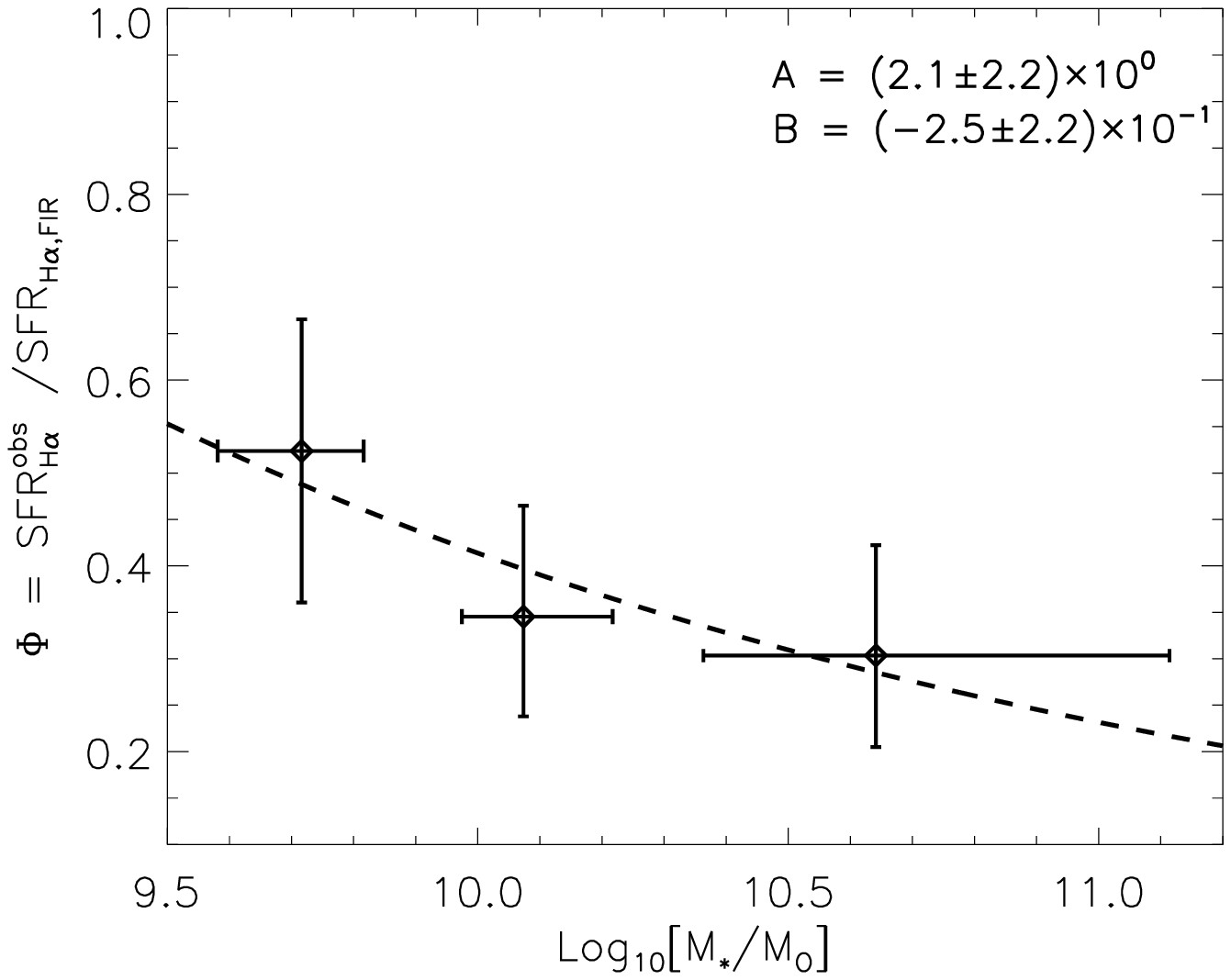}
   \caption{The figures show the parameter $\Phi$, which defines the
     dominant SFR component -- the observed H$\alpha$ or
     L(8--1000\,$\mu$m) luminosity -- as a function of rest-frame
     ($u-z$) colour, observed {\it Spitzer}-3.6\,$\mu$m magnitude and
     stellar mass.  This parametrisation is presented
     \S\,\ref{empirical_SFR} and can be used to aid the common lack of
     the obscured (or unobscured) SFR tracers in high-$z$
     galaxies. The inset values at the top of each figure correspond
     to a simple linear fit, ${\rm Log_{10}}(\Phi)=A+B\,\times\,X$, as
     presented in Eqn.~\ref{eqn_sfr_mass3}.}
   \label{f_fig}
\end{figure}

\subsubsection{An alternative SFR estimator}
\label{empirical_SFR}

In the case when $L({\rm H\alpha})_{\rm obs}\ll a_{\rm FIR}\,
L(8-1000\,\mu{\rm m})$ (see Eqn.~\ref{eqn3}), the far-IR represents
the dominant contribution to the SFR. In the literature, the
transition from a dominant unobscured to a dominant obscured SFR
component has usually been presented as a function of far-IR
luminosity, using H$\alpha$ (e.g.\ \citealt{Villar11}), far-IR
(e.g.\ \citealt{Roseboom12}) or ultra-violet (e.g.\ \citealt{Buat10})
samples. These correlations present large scatter but they all agree
that the brighter the far-IR, the more dominant it becomes as a tracer
of SFR.

Making use of the dependencies shown in
  Fig.~\ref{stack_properties_fig2}, we provide an empirical SFR
  estimate which can be used to alleviate the lack of far-IR
  measurements. We portray this idea in Figure~\ref{f_fig}, where we
show the ratio $\Phi = L({\rm H\alpha})_{\rm obs}/(L({\rm
  H\alpha})_{\rm obs} + a_{\rm FIR}\, L(8-1000\,\mu{\rm m}))$ as a
function of three different parameters. Note that $\Phi$ ranges from 0
to 1 and traces the ratio between unobscured (observed H$\alpha$) and
total (far-IR \& H$\alpha$) SFR. We see that the observed H$\alpha$
luminosity dominates ($>\,$50\%) the total SFR when galaxies have low
stellar masses ($\lesssim$\,10$^{9.8}$\,M$_\odot$), or blue colours
($[u-z]_{\rm rest}\lesssim0.9$), or {\it Spitzer}-3.6\,$\mu$m
photometry fainter than 22\,mag (Vega). For HiZELS at $z=1.47$ (or
similar population), we parametrise the SFR as follows,

\begin{equation}
    {\rm SFR_{H\alpha,\Phi}} = 7.9\times10^{-42}\,
    \frac{L({\rm H}\alpha)_{\rm obs}}{\Phi}
    \label{eqn_sfr_mass1}
\end{equation}


\noindent
where $\Phi$ is obtained empirically by a linear fit,

\begin{equation}
   {\rm Log}_{10}[\Phi] = 
   A + B\times X
   \label{eqn_sfr_mass3}
\end{equation}

Note that $\Phi$ should be lower than unity and
  preferentially higher than 0.15 (see Fig.\ref{f_fig}). We find $A =
0.03\pm0.16$ and $B = -0.35\pm0.15$ if $X$ is the rest-frame ($u-z$)
colour or $A = -5.5\pm1.6$ and $B = 0.236\pm0.071$ if $X$ is the
observed {\it Spitzer}-3.6\,$\mu$m magnitude. The bottom figure shows
a weaker correlation as a function of stellar masses, although
indicates that sources with measured $M_\star$ have, in general, a SFR
dominated by the far-IR emission.

As noted before, these empirical relations can be used to provide a
better proxy for the SFR when far-IR estimates are absent for high-$z$
H$\alpha$-emitting galaxies. This is especially useful for galaxies
presenting high stellar-masses, given that the far-IR luminosity
dominates over the observed H$\alpha$ as a tracer of the SFR. We
remind that these equations are only valid for the range of parameters
discussed in this work, so extrapolations should be taken with
caution.

\subsubsection{A Discrepancy in previous SFR tracers}

We note that in the case when the far-IR emission is by far the
dominant SFR tracer, the parametrisations from the literature
(SFR$_{\rm FIR}$ and SFR$_{\rm H\alpha,FIR}$, see
Table~\ref{sfrs_all_sample}) do not agree. Indeed, the following ratio
does not converge to unity when $L({\rm H\alpha})_{\rm obs}\rightarrow
0$:

$$
  \begin{array}{ll}
    \frac{\rm SFR_{H\alpha,FIR}}{\rm SFR_{FIR}} &
    = \frac{7.9\times10^{-42}[L({\rm H\alpha})_{\rm obs}+2.5\times10^{-3} L(8-1000\,\mu{\rm m})]}
     {4.5\times10^{-44} L(8-1000\,\mu{\rm m})} \\
     & = 0.44
  \end{array} 
$$

This is not related to the assumed IMF, as both estimates use
Salpeter. The 0.44 factor tells us that, if the SFR$_{\rm
  H\alpha,FIR}$ calibration is correct, then the equation for the
far-IR-only calibration assumes that there is a typical additional
contribution of 2.27 times more star-formation associated with an
unobscured component not traced by the far-IR. The accuracy of this
assumption will clearly depend on the nature of the population being
studied. To re-calibrate the SFR estimators is beyond the scope of
this paper, although we stress the need to use the most suitable SFR
tracer, based on the similarity to the parent population used to
define the correlations in the local Universe.

%
\begin{figure}
   \centering
   \includegraphics[scale=0.6]{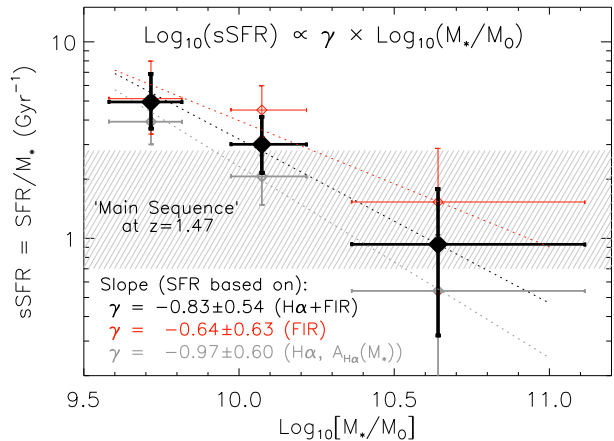}
   \caption{The specific star-formation rate (SFR/$M_\star$) for the
     HiZELS sample at $z=1.47$ as a function of stellar mass. The
     thick black and red symbols show the derived sSFR using SFR$_{\rm
       H\alpha,FIR}$ and SFR$_{\rm FIR}$ respectively (see
     Table~\ref{sfrs_all_sample}). Symbols in light grey are the sSFR
     derived using intrinsic H$\alpha$ luminosities counting for
     extinction using a parametrisation of $A_{\rm H\alpha}$ as a
     function of $M_\star$ (\citealt{bGarn10}; see definition in
     Table~\ref{sfrs_all_sample}). Error bars show the 68\,per~cent
     confidence levels. The shaded line-filled region shows the `main
     sequence' defined in \citet{Elbaz11} at $z=1.5$, corrected by a
     0.15\,dex factor in stellar mass in order to account for the
     TP-AGB component not included in \citeauthor{Elbaz11}'s analysis
     (see \S\,\ref{stellar_mass_section_disc}).}
   \label{sSFR_fig}
\end{figure}

\subsection{Stellar masses and the far-IR power}
\label{stellar_mass_section_disc}

\citet{bGarn10}, using SDSS sources at $\left<z\right>\approx0.08$,
showed that low-mass star-forming galaxies tend to have less
extinction (less obscured star-formation activity) than more massive
ones (see bottom-right panel in Fig.~\ref{stack_properties_fig1}).
Naively speaking, it is expected that as galaxies build up their
stellar masses, they can provide deeper potential wells which could
then help retain their enriched gas (feedback from star-formation
activity) and provide dense environments where colossal numbers of
stars can rapidly form. Under this assumption, we would expect that
massive galaxies would tend to have higher metallicities that can
facilitate the creation of a significant amount of dust which cannot
escape from the galaxy system, implying typically higher optical
depths around star-forming regions. This might be a fundamental reason
to explain the increasing component of heavily extinguished star
formation in more massive galaxies. 

We find that the $A_{\rm H\alpha}(M_\star)$ relation presented by
\citet{bGarn10} seems to have little or no evolution up to $z=1.47$, a
result that confirms previous analyses by \citet{Sobral12} and
\citet{Stott13}. This put important constraints on the Cosmic
evolution of the dust properties up to $z=1.47$, and suggests that the
metal enrichment and the dust covering factors do not evolve
significantly as a function of redshift. We stress that this
correlation is proved using galaxies with $M_\star\gs
10^{9.5}$\,M$_\odot$, so a vast number of galaxies are excluded in
this particular analysis -- especially those biased towards high
specific star-formation rates. Actually, HiZELS galaxies that do have
stellar mass estimates are on average more far-IR luminous than the
others. We show this behaviour in Figure~\ref{f_fig}, where we present
the parameter $\Phi$, which shows the dominant SFR tracer (observed
H$\alpha$ or far-IR luminosity), as a function of rest-frame $(u-z)$
colour, 3.6\,$\mu$m and $M_\star$. We observe that $\Phi$ is always
$\ls\,0.5$, i.e.\ those galaxies with available stellar mass estimates
are in general dominated by obscured star-formation rather than
unextinguished H$\alpha$.

It is important to note that stellar masses are derived from a
SED fit to the rest-frame UV to near-IR photometry, taking into
account five main assumptions: metallicity, age, IMF, reddening and
star-formation history (see \S\,\ref{stellar_mass_section}). The
correlation between this SED-derived SFR and the one derived using the
far-IR luminosity is known to present a large scatter
(\citealt{Buat10, bWijesinghe11}), so stellar masses should be used
with caution, especially for the most obscured galaxies where the UV
could be under-estimating the more heavily obscured star-forming
regions (see \citealt{daCunha08}). We have tried to minimise this
bias by introducing a SED coverage right up to the rest-frame
near-IR bands to make the SED fit not critically dependent upon
rest-frame UV colour (i.e.\ minimising changes in the mass to light
ratio).

The significance of the linear fit's slopes shown in
Figs.~\ref{stack_properties_fig1}\,\&\,\ref{stack_properties_fig2}
suggests that $A_{\rm H\alpha}$ has a slightly stronger correlation
with {\it Spitzer}-3.6\,$\mu$m magnitude than with stellar mass
(Figs.~\ref{stack_properties_fig1}\,\&\,\ref{stack_properties_fig2}).
This is likely to be because, although the Spitzer 3.6\,$\mu$m
magnitude can be used as a broad stellar mass estimator, it also has
some sensitivity to young stars from recent star formation.

\subsection{HiZELS on the main sequence}

In Fig.~\ref{sSFR_fig}, we compare the sSFR using three different
methods: based on the SFR derived from using a combination of
H$\alpha$ and far-IR luminosity (\citealt{Kennicutt09}); based on that
derived using the far-IR only (\citealt{Kennicutt98}); and based on
the optically-derived SFR using $A_{\rm H\alpha}(M_\star)$ (see
definitions in Table~\ref{sfrs_all_sample}). We find that roughly all
sSFR estimates agree, and our preferred SFR$_{\rm H\alpha,FIR}$
estimate unsurprisingly lies between the other two predictions. All
estimators broadly agree within the errors, although slight
discrepancies tend to appear at large stellar masses where the far-IR
becomes the dominant SFR tracer. These results suggest that assuming a
constant SFR, HiZELS galaxies with lower stellar masses
(M$_\star\,\sim\,10^{9.7}\,$M$_\odot$) are five times more efficient
in doubling their stellar mass than more massive ones
(M$_\star\,\sim\,10^{10.7}\,$M$_\odot$), in other words, the
characteristic time that HiZELS galaxies take to double their stellar
mass is $\tau\approx1/5\,$Gyr and $\approx$\,1\,Gyr at low and high
M$_\star$, respectively.

The expected sSFR for a `normal' star-forming galaxy at $z\sim 1.5$ on
the main sequence is 1.4$^{+1.0}_{-0.5}$\,Gyr$^{-1}$. Note that to be
consistent with our analyses, we have decreased \citet{Elbaz11}'s
stellar masses by 0.15\,dex in order to take into account the TP-AGB
component (see \S\,\ref{stellar_mass_section}). This `typical' sSFR is
slightly lower than the typical value measured for the HiZELS sample
(see Fig.~\ref{sSFR_fig}). As such, the HiZELS sample would be classed
as galaxies with a `starburstiness' value of $R_{\rm BS}={\rm
  sSFR/sSFR_{MS}(z)}\approx2$ at $z=1.47$. We note that at this
redshift, the range in stellar population is narrower and the
H$\alpha$ contribution from stars that are not related to star
formation is likely to be negligible (e.g.\ \citealt{Buat10}).

Taking a look at the characteristic break of the H$\alpha$ luminosity
function at $z\approx 1.47$, $L^*_{\rm
  H\alpha}=10^{42.50\pm0.23}$\,erg\,s$^{-1}$ \citep{Sobral12}, we can
confidently say that the HiZELS survey is not identifying the most
extreme and rare galaxies, although $L^*_{\rm H\alpha}$ galaxies
(which dominate the SFR density) tend to fall in the starburst regime
under the \citet{Elbaz11} definition. As starburst galaxies are only a
small fraction in \citeauthor{Elbaz11}'s studies, this might suggest
that typical H$\alpha$ emitters (usually not heavily obscured compared
to a far-IR selected sample; $A_{\rm H\alpha}\ls 2$) are not included
in their work due to selection effects. As most of the HiZELS galaxies
have similar properties to the `normal' ones which define the local
\citet{Kennicutt09}'s SFR relations (i.e.\ allowing reliable SFR
estimates), we suggest the main sequence might be only valid for those
star-forming galaxies which are preferentially obscured
(far-IR-selected). It suggests that we cannot easily compare our
results with previous studies which declare `star-forming' galaxies to
be simply those detected in the far-IR.

It is important to highlight the systematic deviation from the
so-called main sequence for star-forming galaxies (\citealt{Elbaz11})
at low stellar masses. This is mostly due to a selection effect
introduced by the detectability of H$\alpha$ in the HiZELS survey.
Basically, Fig.~\ref{sSFR_fig} does not include the faint H$\alpha$
population composed by a large number of low-mass dwarf galaxies, and
in lower number those which are heavily obscured. The most `starbusty'
ones are selected in this work (especially at small stellar masses),
implying that HiZELS is only sensitive to `main-sequence' galaxies at
high stellar masses ($M_\star\gs10^{10.2}\,M_\odot$).

\section{Conclusions}
\label{conclusion_section}

This work provides for the first time a detailed statistical
description of the far-IR SED for a uniform sample of 443
H$\alpha$-selected star-forming galaxies at $z=1.47$. The sources are
selected from the HiZELS coverage in the UDS and COSMOS fields, and
the measured far-IR properties are obtained from {\it Spitzer} (24,
70\,$\mu$m), {\it Herschel} (100, 160, 250, 350, and 500\,$\mu$m) and
AzTEC (1100\,$\mu$m) images.

We find that the sample of HiZELS galaxies (after removing possible
AGN) have a median far-IR luminosity of $L(8-1000\,\mu {\rm m}) =
10^{11.41^{+0.04}_{-0.06}}\,{\rm L}_\odot$, i.e.\ they are typical
luminous IR galaxies at $z=1.47$ presenting SFR$_{\rm
  H\alpha,FIR}=32\pm5$\,M$_\odot$\,yr$^{-1}$. Our results have been
possible to achieve thanks to a stacking analysis given that these
galaxies are generally beneath the noise levels of present far-IR
images. In particular, only 2\,per cent of the sample is directly
detected by {\it Herschel} at 250\,$\mu$m, a population composed by
massive and heavily obscured galaxies with far-IR luminosities of
$\sim$\,10$^{12.1}$\,L$_\odot$.

We measure a typical H$\alpha$ extinction of $A_{\rm
  H\alpha}=1.0\pm0.2$\,mag for the full HIZELS sample at $z=1.47$, in
excellent agreement with typical extinctions seen locally
(\citealt{bGarn10}) and at high-$z$ (\citealt{Sobral12,Stott13}). We
find little or no evolution up to $z=1.47$ for the correlation between
stellar mass and H$\alpha$ extinction proposed by
\citet{bGarn10}. These results suggest the dust properties do not
change considerably within this redshift range, giving important
constraints on the Cosmic evolution of the dust covering factors and
the properties of metallicity enrichment.

We find that HiZELS galaxies tend to deviate from the `main-sequence'
for star-forming galaxies. This is mostly due to a selection effect
given that only the most starbusty ones are above the H$\alpha$
threshold, especially at low-stellar masses. We find that the
inclusion of far-IR data to obtain better SFR estimates becomes
especially important at high stellar masses.

Our large H$\alpha$ sample has allowed us to explore a large part of
parameter space. In particular, we are able to find a clear
correlation between the far-IR luminosity on rest-frame $u$-$z$ colour
and {\it Spitzer}-3.6\,$\mu$m fluxes. We find that in HiZELS galaxies
presenting red optical $(u-z)_{\rm rest}\,\lesssim\,0.9$ colours or
faint 3.6\,$\mu$m fluxes ($\gtrsim\,22$\,mag, Vega), the observed
H$\alpha$ luminosity becomes the dominant SFR tracer (rather than
far-IR luminosity). We use these dependencies to propose a recipe to
precisely estimate the SFR in cases where far-IR data are absent in
these high redshift galaxies (see Eqn.~\ref{eqn_sfr_mass1}). This
alternative parametrisation is valid for samples with L(${\rm
  H\alpha}$)$_{\rm obs}$/L$(\rm 8-1000\,\mu m)\approx 2/1000$,
i.e.\ when both obscured and unobscured component have similar
contributions to the total SFR.

\section*{Acknowledgements}

We thank the anonymous referee for the useful coments
  that helped improve this paper. E.\ Ibar agradece el financiamento
de CONICYT/ FONDECYT por el proyecto de postdoctorado
N$^\circ$:3130504. DS acknowledges financial support from the
Netherland Organisation for Scientific research (NWO) through a Veni
fellowship. IRS acknowledges support from STFC, a Leverhulme
Fellowship, the ERC Advanced Investigator programme DUSTYGAL and a
Royal Society/Wolfson Merit Award. RJI acknowledges support in the
form of ERC Advanced Investigator programme, {\sc cosmicism}. The
HiZELS data is based on observations obtained using both the Wide
Field Camera on the 3.8-m United Kingdom Infrared Telescope (operated
by the Joint Astronomy Centre on behalf of the Science and Technology
Facilities Council of the U.K.) and Suprime-Cam on the 8.2\,m Subaru
Telescope, which is operated by the National Astronomical Observatory
of Japan.This research has made use data from the HerMES project
(http://hermes.sussex.ac.uk/). HerMES is a {\it Herschel} Key Program
utilising Guaranteed Time from the. SPIRE has been developed by a
consortium of institutes led by Cardiff Univ. (UK) and including:
Univ. Lethbridge (Canada); NAOC (China); CEA, LAM (France); IFSI,
Univ. Padua (Italy); IAC (Spain); Stockholm Observatory (Sweden);
Imperial College London, RAL, UCL-MSSL, UKATC, Univ. Sussex (UK); and
Caltech, JPL, NHSC, Univ. Colorado (USA). This development has been
supported by national funding agencies: CSA (Canada); NAOC (China);
CEA, CNES, CNRS (France); ASI (Italy); MCINN (Spain); SNSB (Sweden);
STFC, UKSA (UK); and NASA (USA). PACS has been developed by a
consortium of institutes led by MPE (Germany) and including UVIE
(Austria); KUL, CSL, IMEC (Belgium); CEA, OAMP (France); MPIA
(Germany); IFSI, OAP/AOT, OAA/CAISMI, LENS, SISSA (Italy); IAC
(Spain). This development has been supported by the funding agencies
BMVIT (Austria), ESA-PRODEX (Belgium), CEA/CNES (France), DLR
(Germany), ASI (Italy), and CICYT/MCYT (Spain). This research has made
use of the NASA/IPAC Infrared Science Archive, which is operated by
the Jet Propulsion Laboratory, California Institute of Technology,
under contract with the National Aeronautics and Space
Administration. This work is based in part on observations made with
the Spitzer Space Telescope, which is operated by the Jet Propulsion
Laboratory, California Institute of Technology under a contract with
NASA.

\bibliographystyle{mn2e}
\bibliography{paper-2013_v15_astro-ph.bib}

%
\begin{landscape}
\begin{table}
 \caption{The stacked far-IR fluxes found for the different selection
   criteria presented throughout this HiZELS study. `Ref' stands for a
   reference number (see Table~\ref{table2}). In the column called
   `Field': U\,$=$\,UDS and C\,$=$\,COSMOS. `N' corresponds to the
   total number of sources in the HiZELS sample satisfying the
   `Selection criterion'. Details on the flux density measures and
   Monte-Carlo-based errors are described in
   (\S\,\ref{SED_stack_section}). Flux densities and errors are
   corrected by the empirical calibration factor $\eta$ (see
   \S\,\ref{empirical_cal_section}). If fluxes are zero, then the
   measured flux was found to be negative.}
\begin{tabular}{|cccccccccccc|}
\hline
Ref. &
Field & 
Selection criterion &
$N$ &
$S_{\rm 24\mu m}$ &
$S_{\rm 70\mu m}$ &
$S_{\rm 100\mu m}$ &
$S_{\rm 160\mu m}$ &
$S_{\rm 250\mu m}$ &
$S_{\rm 350\mu m}$ &
$S_{\rm 500\mu m}$ &
$S_{\rm 1100\mu m}$ \\
 &
 & 
 &
 &
(mJy) &
(mJy) &
(mJy) &
(mJy) &
(mJy) &
(mJy) &
(mJy) &
(mJy) \\
\hline
 1 & C & ${\rm S(H\alpha)>6\times10^{-17} \,\, \&\,\,  S([OII])>1.3\times10^{-17}}$ & 113 & 0.05$\pm$0.02 & 
0.52$\pm$0.23 & 
1.97$\pm$0.51 & 
1.66$\pm$0.89 & 
3.89$\pm$0.62 & 
4.46$\pm$0.78 & 
3.68$\pm$0.77 & 
0.25$\pm$0.16 \\ 
 2 & U & ${\rm S(H\alpha)>6\times10^{-17} \,\, \&\,\,   S([OII])>1.3\times10^{-17}}$ & 123 & 0.06$\pm$0.02 & 
0.40$\pm$0.39 & 
0.63$\pm$0.67 & 
0.89$\pm$1.25 & 
4.70$\pm$0.73 & 
4.92$\pm$0.82 & 
3.00$\pm$0.78 & 
0.42$\pm$0.18 \\ 
 3 & U+C & ${\rm all}$ & 443 & 0.04$\pm$0.01 & 
0.39$\pm$0.14 & 
1.19$\pm$0.32 & 
1.31$\pm$0.53 & 
3.44$\pm$0.43 & 
3.52$\pm$0.48 & 
2.40$\pm$0.42 & 
0.22$\pm$0.10 \\ 
 4 & U+C & ${\rm 41.6 < Log_{10}[L(H\alpha)_{obs}] \leq 42.1}$ & 129 & 0.04$\pm$0.01 & 
0.46$\pm$0.24 & 
1.31$\pm$0.43 & 
0.88$\pm$1.03 & 
3.20$\pm$0.59 & 
3.85$\pm$0.74 & 
2.26$\pm$0.72 & 
0.22$\pm$0.16 \\ 
 5 & U+C & ${\rm 42.1 < Log_{10}[L(H\alpha)_{obs}] \leq 42.25}$ & 133 & 0.05$\pm$0.02 & 
0.31$\pm$0.26 & 
1.64$\pm$0.57 & 
1.27$\pm$0.97 & 
4.12$\pm$0.63 & 
4.18$\pm$0.73 & 
2.95$\pm$0.69 & 
0.26$\pm$0.17 \\ 
 6 & U+C & ${\rm 42.25 < Log_{10}[L(H\alpha)_{obs}] \leq 43.0}$ & 181 & 0.05$\pm$0.02 & 
0.47$\pm$0.23 & 
1.15$\pm$0.40 & 
2.17$\pm$0.84 & 
3.17$\pm$0.51 & 
2.76$\pm$0.57 & 
2.01$\pm$0.57 & 
0.22$\pm$0.17 \\ 
 7 & U+C & ${\rm 41.6 < Log_{10}[L(H\alpha)_{\rm int}([OII]/{\rm H\alpha})] \leq 42.4}$ &  85 & 0.04$\pm$0.02 & 
0.26$\pm$0.32 & 
1.23$\pm$0.58 & 
0.00$\pm$1.24 & 
2.90$\pm$0.69 & 
3.31$\pm$0.85 & 
2.33$\pm$0.89 & 
0.27$\pm$0.20 \\ 
 8 & U+C & ${\rm 42.4 < Log_{10}[L(H\alpha)_{\rm int}([OII]/{\rm H\alpha})] \leq 42.6}$ &  69 & 0.04$\pm$0.02 & 
0.43$\pm$0.39 & 
1.28$\pm$0.73 & 
0.00$\pm$1.52 & 
5.17$\pm$0.86 & 
6.11$\pm$1.05 & 
4.25$\pm$1.00 & 
0.33$\pm$0.24 \\ 
 9 & U+C & ${\rm 42.6 < Log_{10}[L(H\alpha)_{\rm int}([OII]/{\rm H\alpha})] \leq 44.1}$ & 121 & 0.07$\pm$0.02 & 
0.54$\pm$0.30 & 
1.98$\pm$0.56 & 
3.31$\pm$1.02 & 
4.23$\pm$0.66 & 
4.17$\pm$0.75 & 
2.95$\pm$0.73 & 
0.07$\pm$0.18 \\ 
 10 & U+C & ${\rm 41.6 < Log_{10}[L(H\alpha)_{\rm int}(M_\star)] \leq 42.4}$ & 151 & 0.02$\pm$0.01 & 
0.04$\pm$0.23 & 
0.85$\pm$0.33 & 
0.00$\pm$1.08 & 
1.38$\pm$0.48 & 
1.68$\pm$0.59 & 
0.92$\pm$0.61 & 
0.08$\pm$0.15 \\ 
 11 & U+C & ${\rm 42.4 < Log_{10}[L(H\alpha)_{\rm int}(M_\star)] \leq 42.6}$ & 127 & 0.04$\pm$0.01 & 
0.29$\pm$0.27 & 
0.35$\pm$0.46 & 
0.66$\pm$0.98 & 
2.85$\pm$0.57 & 
2.73$\pm$0.67 & 
1.98$\pm$0.68 & 
0.29$\pm$0.19 \\ 
 12 & U+C & ${\rm 42.6 < Log_{10}[L(H\alpha)_{\rm int}(M_\star)] \leq 44.1}$ & 162 & 0.10$\pm$0.03 & 
0.91$\pm$0.24 & 
1.99$\pm$0.43 & 
2.89$\pm$0.79 & 
6.16$\pm$0.75 & 
6.20$\pm$0.81 & 
4.47$\pm$0.71 & 
0.40$\pm$0.17 \\ 
 13 & U+C & ${\rm 41.6 < Log_{10}[L(H\alpha)_{\rm int}([u-z]_{\rm rest})] \leq 42.4}$ & 153 & 0.04$\pm$0.01 & 
0.33$\pm$0.23 & 
1.14$\pm$0.41 & 
0.00$\pm$0.88 & 
2.54$\pm$0.52 & 
2.98$\pm$0.64 & 
1.79$\pm$0.65 & 
0.14$\pm$0.15 \\ 
 14 & U+C & ${\rm 42.4 < Log_{10}[L(H\alpha)_{\rm int}([u-z]_{\rm rest})] \leq 42.6}$ & 172 & 0.04$\pm$0.01 & 
0.41$\pm$0.23 & 
0.93$\pm$0.42 & 
1.74$\pm$0.75 & 
3.60$\pm$0.55 & 
3.53$\pm$0.62 & 
2.48$\pm$0.60 & 
0.22$\pm$0.16 \\ 
 15 & U+C & ${\rm 42.6 < Log_{10}[L(H\alpha)_{\rm int}([u-z]_{\rm rest})] \leq 44.1}$ & 115 & 0.08$\pm$0.02 & 
0.54$\pm$0.27 & 
1.58$\pm$0.48 & 
2.33$\pm$0.92 & 
4.36$\pm$0.66 & 
4.41$\pm$0.77 & 
3.36$\pm$0.74 & 
0.41$\pm$0.21 \\ 
 16 & U+C & ${\rm 9.5 < Log_{10}[M_*/M_\odot] \leq 9.9}$ &  90 & 0.03$\pm$0.01 & 
0.22$\pm$0.30 & 
0.30$\pm$0.57 & 
0.01$\pm$1.16 & 
2.04$\pm$0.62 & 
1.74$\pm$0.75 & 
1.02$\pm$0.78 & 
0.00$\pm$0.21 \\ 
 17 & U+C & ${\rm 9.9 < Log_{10}[M_*/M_\odot] \leq 10.3}$ & 443 & 0.06$\pm$0.02 & 
0.66$\pm$0.27 & 
1.52$\pm$0.47 & 
1.43$\pm$0.87 & 
4.37$\pm$0.59 & 
4.37$\pm$0.65 & 
2.60$\pm$0.58 & 
0.21$\pm$0.15 \\ 
 18 & U+C & ${\rm 10.3 < Log_{10}[M_*/M_\odot] \leq 11.8}$ & 443 & 0.07$\pm$0.02 & 
0.62$\pm$0.22 & 
1.68$\pm$0.41 & 
2.39$\pm$0.73 & 
5.20$\pm$0.63 & 
5.73$\pm$0.72 & 
3.91$\pm$0.61 & 
0.43$\pm$0.14 \\ 
 19 & U+C & ${\rm -1.20 < Log_{10}[[OII]/H\alpha] \leq -0.50}$ &  84 & 0.05$\pm$0.02 & 
0.42$\pm$0.36 & 
2.09$\pm$0.67 & 
2.16$\pm$1.16 & 
3.45$\pm$0.70 & 
3.27$\pm$0.82 & 
2.28$\pm$0.83 & 
0.02$\pm$0.22 \\ 
 20 & U+C & ${\rm -0.50 < Log_{10}[[OII]/H\alpha] \leq -0.25}$ &  90 & 0.07$\pm$0.02 & 
0.42$\pm$0.31 & 
1.81$\pm$0.59 & 
1.04$\pm$1.23 & 
5.78$\pm$0.82 & 
5.51$\pm$0.91 & 
3.62$\pm$0.85 & 
0.38$\pm$0.20 \\ 
 21 & U+C & ${\rm -0.25 < Log_{10}[[OII]/H\alpha] \leq 0.40}$ & 101 & 0.04$\pm$0.01 & 
0.37$\pm$0.31 & 
1.43$\pm$0.55 & 
0.20$\pm$0.97 & 
3.27$\pm$0.67 & 
3.51$\pm$0.80 & 
2.81$\pm$0.83 & 
0.22$\pm$0.19 \\ 
 22 & U+C & $-0.5 < (u-z)_{\rm rest} \leq 0.9$ & 146 & 0.02$\pm$0.01 & 
0.25$\pm$0.26 & 
0.11$\pm$0.45 & 
0.00$\pm$0.91 & 
1.32$\pm$0.47 & 
1.32$\pm$0.58 & 
1.06$\pm$0.61 & 
0.33$\pm$0.18 \\ 
 23 & U+C & $0.9 < (u-z)_{\rm rest} \leq 1.5$ & 153 & 0.05$\pm$0.02 & 
0.43$\pm$0.23 & 
1.30$\pm$0.46 & 
1.98$\pm$0.85 & 
3.01$\pm$0.54 & 
2.86$\pm$0.63 & 
1.59$\pm$0.61 & 
0.00$\pm$0.16 \\ 
 24 & U+C & $1.5 < (u-z)_{\rm rest} \leq 3.0$ & 139 & 0.10$\pm$0.03 & 
0.71$\pm$0.25 & 
2.19$\pm$0.51 & 
2.47$\pm$0.89 & 
6.87$\pm$0.83 & 
7.28$\pm$0.93 & 
5.09$\pm$0.80 & 
0.44$\pm$0.17 \\ 
 25 & U+C & $19.0 < 3.6\mu{\rm m} \leq 21.1$ & 117 & 0.12$\pm$0.04 & 
0.91$\pm$0.29 & 
1.93$\pm$0.58 & 
3.82$\pm$1.00 & 
8.47$\pm$0.99 & 
8.76$\pm$1.08 & 
5.84$\pm$0.90 & 
0.46$\pm$0.19 \\ 
 26 & U+C & $21.1 < 3.6\mu{\rm m} \leq 22.1$ & 142 & 0.05$\pm$0.02 & 
0.35$\pm$0.24 & 
1.47$\pm$0.48 & 
1.62$\pm$0.92 & 
3.28$\pm$0.57 & 
3.17$\pm$0.66 & 
1.87$\pm$0.63 & 
0.20$\pm$0.17 \\ 
 27 & U+C & $22.1 < 3.6\mu{\rm m} \leq 25.0$ & 136 & 0.01$\pm$0.01 & 
0.00$\pm$0.24 & 
0.03$\pm$0.43 & 
0.00$\pm$0.79 & 
0.00$\pm$0.46 & 
0.40$\pm$0.58 & 
0.59$\pm$0.62 & 
0.00$\pm$0.17 \\ 
 28 & U+C & ${\rm AGN}$ &  70 & 0.07$\pm$0.02 & 
0.61$\pm$0.36 & 
1.31$\pm$0.66 & 
4.90$\pm$1.36 & 
3.96$\pm$0.77 & 
3.60$\pm$0.89 & 
2.63$\pm$0.90 & 
0.00$\pm$0.24 \\ 
\hline
\end{tabular}
\label{table5}
\end{table}
\end{landscape}

%
\begin{table*}
 \caption{The derived far-IR properties found for the different
   selection criteria (`Ref.' are shown in Table~\ref{table5})
   presented throughout this HiZELS study. The median and the 68\%
   distribution of the observed H$\alpha$ luminosity is presented in
   the second column. The fitting procedure used to extract the far-IR
   properties is explained in \S\ref{SED_fitting_section}.}
\begin{tabular}{|ccccccc|}
\hline
Ref. & 
${\rm Log_{10}[L(H\alpha)_{obs}]}$ &
${\rm Log}_{10}[L({\rm 8-1000\mu m})/L_\odot]$ &
${\rm SFR_{\rm H\alpha,FIR}}$ &
$A_{\rm H\alpha}$ &
$T_{\rm dust}$ &
$\alpha_{\operatorname{mid-IR}}$ \\
 & 
(68\,\% distribution) &
 &
${\rm (M_\odot\, yr^{-1})}$ &
$({\rm mag})$ &
(K) & \\
\hline
 1 &
$42.23^{+0.18}_{-0.19}$ &
$11.48^{+0.07}_{-0.06}$ &
$36.2^{+ 6.9}_{- 6.3}$ &
$1.1^{+0.2}_{-0.2}$ &
$23^{+ 2}_{- 2}$ &
$ 1.9^{+ 0.5}_{- 0.2}$ \\
 2 &
$42.22^{+0.13}_{-0.17}$ &
$11.48^{+0.08}_{-0.08}$ &
$36.2^{+ 7.3}_{- 6.5}$ &
$1.1^{+0.2}_{-0.2}$ &
$23^{+ 3}_{- 2}$ &
$ 2.1^{+ 0.6}_{- 0.3}$ \\
 3 &
$42.22^{+0.15}_{-0.22}$ &
$11.41^{+0.04}_{-0.06}$ &
$32.1^{+ 5.4}_{- 5.0}$ &
$1.0^{+0.2}_{-0.2}$ &
$24^{+ 1}_{- 1}$ &
$ 1.9^{+ 0.3}_{- 0.2}$ \\
 4 &
$41.99^{+0.08}_{-0.15}$ &
$11.39^{+0.07}_{-0.10}$ &
$25.9^{+ 5.8}_{- 5.1}$ &
$1.3^{+0.2}_{-0.2}$ &
$24^{+ 2}_{- 2}$ &
$ 2.0^{+ 0.4}_{- 0.3}$ \\
 5 &
$42.18^{+0.06}_{-0.05}$ &
$11.45^{+0.06}_{-0.07}$ &
$33.2^{+ 6.6}_{- 5.8}$ &
$1.1^{+0.2}_{-0.2}$ &
$24^{+ 3}_{- 2}$ &
$ 2.1^{+ 0.4}_{- 0.4}$ \\
 6 &
$42.35^{+0.13}_{-0.07}$ &
$11.39^{+0.06}_{-0.07}$ &
$36.2^{+ 5.8}_{- 5.2}$ &
$0.8^{+0.2}_{-0.2}$ &
$28^{+ 3}_{- 3}$ &
$ 2.0^{+ 0.4}_{- 0.2}$ \\
 7 &
$42.02^{+0.13}_{-0.16}$ &
$11.36^{+0.07}_{-0.14}$ &
$24.8^{+ 5.9}_{- 5.1}$ &
$1.2^{+0.2}_{-0.3}$ &
$24^{+ 6}_{- 3}$ &
$ 2.0^{+ 0.6}_{- 0.4}$ \\
 8 &
$42.18^{+0.10}_{-0.12}$ &
$11.53^{+0.06}_{-0.08}$ &
$37.7^{+ 7.9}_{- 7.0}$ &
$1.2^{+0.2}_{-0.2}$ &
$23^{+ 2}_{- 2}$ &
$ 2.3^{+ 0.4}_{- 0.3}$ \\
 9 &
$42.30^{+0.16}_{-0.16}$ &
$11.55^{+0.06}_{-0.07}$ &
$42.3^{+ 7.8}_{- 7.1}$ &
$1.1^{+0.2}_{-0.2}$ &
$27^{+ 3}_{- 3}$ &
$ 2.0^{+ 0.3}_{- 0.3}$ \\
 10 &
$42.05^{+0.12}_{-0.19}$ &
$11.12^{+0.13}_{-0.14}$ &
$18.7^{+ 4.9}_{- 3.4}$ &
$0.8^{+0.3}_{-0.2}$ &
$29^{+ 8}_{- 8}$ &
$ 2.0^{+ 0.6}_{- 0.5}$ \\
 11 &
$42.26^{+0.08}_{-0.14}$ &
$11.25^{+0.10}_{-0.09}$ &
$27.9^{+ 5.0}_{- 3.9}$ &
$0.7^{+0.2}_{-0.2}$ &
$24^{+ 4}_{- 3}$ &
$ 2.2^{+ 0.4}_{- 0.3}$ \\
 12 &
$42.34^{+0.14}_{-0.16}$ &
$11.69^{+0.04}_{-0.05}$ &
$53.9^{+10.0}_{- 9.3}$ &
$1.2^{+0.2}_{-0.2}$ &
$24^{+ 1}_{- 1}$ &
$ 1.7^{+ 0.3}_{- 0.2}$ \\
 13 &
$42.04^{+0.09}_{-0.17}$ &
$11.31^{+0.10}_{-0.07}$ &
$24.4^{+ 5.3}_{- 4.6}$ &
$1.1^{+0.2}_{-0.2}$ &
$26^{+ 6}_{- 4}$ &
$ 2.0^{+ 0.4}_{- 0.4}$ \\
 14 &
$42.24^{+0.08}_{-0.09}$ &
$11.40^{+0.06}_{-0.07}$ &
$32.9^{+ 5.6}_{- 5.0}$ &
$0.9^{+0.2}_{-0.2}$ &
$25^{+ 3}_{- 2}$ &
$ 2.0^{+ 0.4}_{- 0.2}$ \\
 15 &
$42.38^{+0.15}_{-0.13}$ &
$11.54^{+0.05}_{-0.08}$ &
$44.9^{+ 7.7}_{- 7.0}$ &
$0.9^{+0.2}_{-0.2}$ &
$24^{+ 2}_{- 1}$ &
$ 1.8^{+ 0.3}_{- 0.3}$ \\
 16 &
$42.23^{+0.12}_{-0.25}$ &
$11.18^{+0.16}_{-0.15}$ &
$25.1^{+ 6.2}_{- 4.0}$ &
$0.6^{+0.2}_{-0.2}$ &
$28^{+ 9}_{- 8}$ &
$ 2.2^{+ 0.5}_{- 0.6}$ \\
 17 &
$42.22^{+0.15}_{-0.22}$ &
$11.51^{+0.04}_{-0.05}$ &
$37.4^{+ 6.9}_{- 6.3}$ &
$1.1^{+0.2}_{-0.2}$ &
$26^{+ 1}_{- 2}$ &
$ 1.9^{+ 0.4}_{- 0.2}$ \\
 18 &
$42.22^{+0.15}_{-0.22}$ &
$11.59^{+0.04}_{-0.06}$ &
$42.6^{+ 8.1}_{- 7.6}$ &
$1.3^{+0.2}_{-0.2}$ &
$23^{+ 1}_{- 1}$ &
$ 1.8^{+ 0.2}_{- 0.2}$ \\
 19 &
$42.18^{+0.22}_{-0.14}$ &
$11.46^{+0.07}_{-0.08}$ &
$33.8^{+ 7.1}_{- 6.3}$ &
$1.1^{+0.2}_{-0.2}$ &
$28^{+ 4}_{- 4}$ &
$ 2.0^{+ 0.6}_{- 0.3}$ \\
 20 &
$42.21^{+0.15}_{-0.21}$ &
$11.59^{+0.07}_{-0.06}$ &
$42.6^{+ 8.8}_{- 7.9}$ &
$1.3^{+0.2}_{-0.2}$ &
$25^{+ 4}_{- 1}$ &
$ 2.2^{+ 0.4}_{- 0.4}$ \\
 21 &
$42.14^{+0.17}_{-0.22}$ &
$11.39^{+0.09}_{-0.07}$ &
$29.7^{+ 6.3}_{- 5.3}$ &
$1.1^{+0.2}_{-0.2}$ &
$25^{+ 4}_{- 4}$ &
$ 2.0^{+ 0.5}_{- 0.3}$ \\
 22 &
$42.23^{+0.16}_{-0.23}$ &
$10.98^{+0.16}_{-0.13}$ &
$20.8^{+ 3.6}_{- 2.5}$ &
$0.5^{+0.2}_{-0.1}$ &
$22^{+11}_{- 4}$ &
$ 2.0^{+ 0.6}_{- 0.4}$ \\
 23 &
$42.22^{+0.16}_{-0.21}$ &
$11.40^{+0.07}_{-0.10}$ &
$32.0^{+ 6.0}_{- 5.4}$ &
$1.0^{+0.2}_{-0.2}$ &
$29^{+ 4}_{- 4}$ &
$ 2.0^{+ 0.6}_{- 0.3}$ \\
 24 &
$42.19^{+0.15}_{-0.17}$ &
$11.70^{+0.04}_{-0.07}$ &
$49.9^{+10.4}_{- 9.8}$ &
$1.5^{+0.2}_{-0.2}$ &
$23^{+ 1}_{- 1}$ &
$ 1.8^{+ 0.3}_{- 0.2}$ \\
 25 &
$42.20^{+0.16}_{-0.17}$ &
$11.77^{+0.04}_{-0.05}$ &
$56.9^{+12.2}_{-11.3}$ &
$1.6^{+0.2}_{-0.2}$ &
$23^{+ 1}_{- 1}$ &
$ 1.8^{+ 0.3}_{- 0.2}$ \\
 26 &
$42.22^{+0.15}_{-0.23}$ &
$11.41^{+0.07}_{-0.08}$ &
$32.4^{+ 5.9}_{- 5.4}$ &
$1.0^{+0.2}_{-0.2}$ &
$26^{+ 3}_{- 2}$ &
$ 2.0^{+ 0.6}_{- 0.3}$ \\
 27 &
$42.20^{+0.15}_{-0.25}$ &
$10.64^{+0.24}_{-0.30}$ &
$16.3^{+ 3.0}_{- 1.9}$ &
$0.2^{+0.2}_{-0.1}$ &
$27^{+13}_{-11}$ &
$ 1.8^{+ 0.7}_{- 0.5}$ \\
 28 &
$42.22^{+0.21}_{-0.22}$ &
$11.56^{+0.07}_{-0.09}$ &
$40.2^{+ 8.8}_{- 7.5}$ &
$1.2^{+0.2}_{-0.2}$ &
$31^{+ 4}_{- 4}$ &
$ 2.2^{+ 0.4}_{- 0.4}$ \\
\hline
\end{tabular}
\label{table2}
\end{table*}

\bsp
\label{lastpage}

\end{document}